\documentclass[universe,article,accept,pdftex,moreauthors]{Definitions/mdpi} 

\firstpage{1} 
\pubvolume{1}
\issuenum{1}
\articlenumber{0}
\pubyear{2025}
\copyrightyear{2025}
\datereceived{23 January 2025} 
\daterevised{22 February 2025} 
\dateaccepted{26 February 2025} 
\datepublished{ } 
\hreflink{https://\\doi.org/} 

\Title{Unveiling the Dynamics in Galaxy Clusters: The Hidden Role of Low-Luminosity Galaxies in Coma}

\TitleCitation{Unveiling the Dynamics in Galaxy Clusters: The Hidden Role of Low-Luminosity Galaxies in Coma}

\Author{Alisson 
 P. Costa $^{\dagger,}$*, Andre. L. B. Ribeiro 
 $^{\dagger,}$*, Flavio R. de M. Neto $^{\dagger}$ and Juarez dos S. Junior
}

\AuthorNames{Alisson Costa, Andre Ribeiro, Flavio Morais-Neto and Juarez dos Santos}

\AuthorCitation{Costa, A.P.; 
 Ribeiro, A.L.B.; de Morais Neto, F.R.; dos Santos Junior, J.}

\address[1]{Laboratorio de Astrofisica Teorica e Observacional, Universidade Estadual de Santa Cruz, Ilhéus 45662-900, Brazil; 
\\} 

\corres{\hangafter=1 \hangindent=1.05em \hspace{-0.82em}Correspondence: alisson.costa@enova.educacao.ba.gov.br (A.P.C.); albr@uesc.br (A.L.B.R.)}

\firstnote{\hangafter=1 \hangindent=1.05em \hspace{-0.82em} These 
 authors contributed equally to this work.}  

\abstract{In this work, we study the Coma cluster, one of the richest and most well-known systems at low redshifts, to explore the importance of low-flux objects in the identification of cluster substructures. In addition, we conduct a study of the infall flow around Coma, considering the presence or absence of low-flux objects across the projected phase space of the cluster. Our results indicate that low-luminosity galaxies play a fundamental role in understanding the dynamical state of galaxy clusters. These galaxies, often overlooked because of their faint nature, serve as sensitive tracers of substructure dynamics and provide crucial insights into the cluster's evolutionary history. We show that not considering the low-flux objects present in clusters can lead to significant underestimates of the numbers of substructures, both in most central parts, in the infall regions, and beyond, connecting to the large-scale structure up to a distance of $\sim$8$\;R_{200}$ from the center of Coma. 
}


\begin{document}

\section{Introduction}
One of the main aspects related to the dynamics in galaxy clusters is the presence of substructures, which are fundamental to understanding the formation of large-scale structures in the $\Lambda$CDM Universe. These substructures not only provide insight into the process of mass accretion, cluster dynamics, and the evolutionary history of galaxies within these clusters but also enhance our understanding of galaxy formation (e.g.,~\cite{2004MNRAS.348..333D, 2018MNRAS.475..853O}).
However, identifying substructures in the Universe is challenging because of several factors, including projection effects and limited observational depth. In this context, the challenge of probing deeper into the cluster samples arises from instrumental limitations. As a result, most dynamical studies of galaxy clusters focus on the brightest objects, those with the highest flux—because they can be detected over larger distances, thereby reducing incompleteness and bias in magnitude-based samples~\cite{2008MNRAS.387.1253W, 2014ApJ...782...23S, mason2023brightest}.
Although this practice is common since bright galaxies tend to be the most massive and often occupy central or dominant positions in clusters and groups, relying solely on bright galaxies for structural analysis may lead to incorrect diagnoses about the true dynamical state of the clusters. Low-flux galaxies often trace more dynamic or disturbed populations 
(e.g.,~\cite{kaviraj2011coincidence, 2018MNRAS.473L..31C}). Thus, dwarf or low-flux galaxies are crucial for tracing the hierarchical formation process of the cluster, and their exclusion from analyses can obscure evidence of recent merger or accretion events.


Some works address this important topic. For instance,
to investigate the environmental dependence of the properties of the galaxy by analyzing how the local density and richness of galaxy systems influence star formation rates and galaxy morphologies, ref.
~\cite{tanaka2004environmental} selected a set of galaxies in the local Universe $0.030 < z < 0.065$, dividing the sample into bright $M_{r} < M_{r}^{*}+1$ and low-luminosity galaxies (also called faint, $M_{r}^{*}+1 < M_{r} < M_{r}^{*}+2$), for a cosmology with $M_{r}^{*}=-21.4$. The study showed that for faint galaxies, there is a critical density of $\log \Sigma_{\text{crit}} \sim 0.4$ galaxies $h^{2}_{75}$$~$$\text{Mpc}^{-2}$, where star formation and morphology change drastically. That is, faint galaxies show more active star formation than bright galaxies for a $B/T$ ratio $<0.2$ (bulge-to-total luminosity ratio). This trend is particularly prominent in low-density regions.

In connection with the previous study, that is, probing the dynamical implications of low-luminosity galaxies in the environments they occupy, ref. ~\cite{2012ApJ...754...98B} performed direct measurements of the oxygen abundance in a set of galaxies with spectroscopy observed by the MMT telescope. Selecting 42 low-luminosity galaxies in the \textit{Spitzer}
 LVL (Local Volume Legacy) survey, with B-band magnitudes that span the range $-10.8 \geq M_{B} \geq -18.8$, the authors investigated the relationships between luminosity and metallicity, and between mass and metallicity. It was found that the relationships involving metallicity suffer from large variations, owing to low statistics for low-luminosity regimes. 
This may introduce biases in metallicity studies associated with mass or luminosity distribution, as these quantities are correlated. This relationship, when well-calibrated, allows low-luminosity galaxies to become more effective tracers of the cluster's substructures, contributing to a more complete understanding of the system's internal dynamics.

These findings emphasize the important role that low-luminosity galaxies play in detecting environmental impacts in the dynamics of clusters. Their heightened sensitivity to local density, marked by significant changes in star formation and morphology, underscores their value in understanding dynamical processes. The absence of these galaxies in analyses could bias results, as it would mask the nuanced interplay between environment and galaxy evolution, particularly in low-density regions where these faint galaxies actively contribute to the overall star formation budget and the dynamical state of the system.
From this perspective, we will revisit the Coma cluster using recently developed techniques to explore the role of low-luminosity galaxies. However, it is crucial to consider the following question: \textit{What significant findings do we already know about Coma?} Since its pivotal use by Fritz Zwicky to shed light on the existence of dark matter (through the inconsistency of the velocity dispersion of galaxies), Coma has been the prototype for several studies on its global dynamics. 

In this context, ref. ~\cite{1987ApJ...317..653F} initially identified only two structures in the central region of Coma using maximum likelihood, indicating that the central region had not yet reached equilibrium.
Ref. ~\cite{1997ApJ...488..136G} also analyzed the central Coma region using a set of 320 galaxies. By applying a 3-D wavelet transform combined with segmentation analysis, they identified a highly fragmented structure with multiple substructures. However, they claim that the statistical reliability of these substructures requires further investigation, as the luminosity limits of the sample were not considered. In~\cite{2005A&A...443...17A}, the authors studied Coma formation by analyzing falling substructures around the cluster. They found a total of 17 substructures in a sample of 920 galaxies, limited to $ R = 19.5$, to avoid incompleteness of the sample. Using a hierarchical approach combined with X-ray data, they identified substructures associated with notable galaxies such as NGC 4874 and NGC 4889 in the nuclear region of Coma.

At the same time, comparing substructures in galaxy clusters as identified in X-rays and optical observations is crucial for understanding the physical processes governing cluster formation, evolution, and dynamics.
Recently, ref. ~\cite{2021A&A...651A..41C} examined the turbulent life of Coma beyond $R_{500}$ using X-ray data from SRG/eROSITA. Their findings revealed a faint X-ray bridge connecting one of the main substructures of Coma, led by NGC 4839, to the cluster core. This result suggests that the NGC 4839 group has already traversed the main cluster, as indicated by two distinct gas shocks in the Coma core. The first shock occurred during NGC 4839's initial passage through the cluster several billion years ago, while the second, described as a ``mini-accretion shock'', is associated with the gas settling back into a quasi-hydrostatic equilibrium in the core.

In this work, we study the Coma cluster, one of the richest and most well-known systems at low redshifts, to explore the importance of low-flux objects in the identification of cluster substructures. In addition, we conduct a study of the infall flow around Coma, considering the presence or absence of low-flux objects. The study aims to reveal possible biases in the dynamic characterization of a cluster, depending on the magnitude or stellar mass limit of the galaxies selected as members.
Throughout the paper, we assume a flat $\Lambda$CDM model with $\Omega_m = 0.3$, $\Omega_\Lambda = 0.7$, $H_0 = 70 \, \mathrm{km/s/Mpc}$. 


\section{Data and Methodology}

\subsection{Data}
Data were obtained from the Sloan Digital Sky Survey (SDSS), Data Release 15 (DR15). Objects were selected within a radius of 550 arcminutes from the center, which is located approximately 194.95$^\circ$ in right ascension and 27.97$^\circ$ in declination, corresponding to the peak of the X-ray emission as observed by the \textit{Chandra X-ray Observatory} e.g. Vikhlinin+2001. Furthermore, the selection was made within a velocity range of 4500 km/s around the cluster velocity, $V=6925$ km/s (or $z=0.0231)$. Membership was determined using the \textit{shiftgapper} method, which involves iteratively grouping galaxies based on their velocity and position \citep{Fadda1996,girardi1996velocity}. This method effectively identifies cluster members by removing interlopers, using gaps in the velocity distribution to distinguish between cluster members and field galaxies in multiple radial bins.
This approach is particularly effective in the outer regions of galaxy clusters, where interlopers (field galaxies or those from other clusters) can contaminate the sample. The method can be summarized as follows:

\textit{1. Radial division into bins:} The cluster is divided into concentric rings (bins) centered on the cluster. Within each bin, all galaxies are considered, and their radial velocity distributions are analyzed.
In this work, we define a 0.6 Mpc radial bin \citep{lopes2009nosocs}.

\textit{2. Variable gap criterion:} Galaxies are first sorted in ascending order of their line-of-sight velocity $v_i$. The velocity gaps are then computed as:
$g_i = v_{i+1} - v_i,~\text{for}~i = 1, 2, ..., N-1$.
To determine whether a galaxy is an interloper, a threshold is defined based on the velocity dispersion of the cluster at each bin. The velocity dispersion $\sigma_v(R)$ is also calculated in the radial bins, and the gap threshold is given by $g_i > 2.5 \times \sigma_v(R)$ \citep{adami1998eso,lopes2009nosocs}.

\textit{3. Iterative process along the radial distance:} The procedure is repeated across all bins, excluding galaxies rejected in previous iterations, until the cluster member count stabilizes.

\textit{4. Independence from dynamical assumptions:} A key strength of the method is its reliance on robust observational criteria rather than specific dynamical models, such as cluster \linebreak mass profiles.

\begin{figure}[H]
\includegraphics[width=10.5 cm]{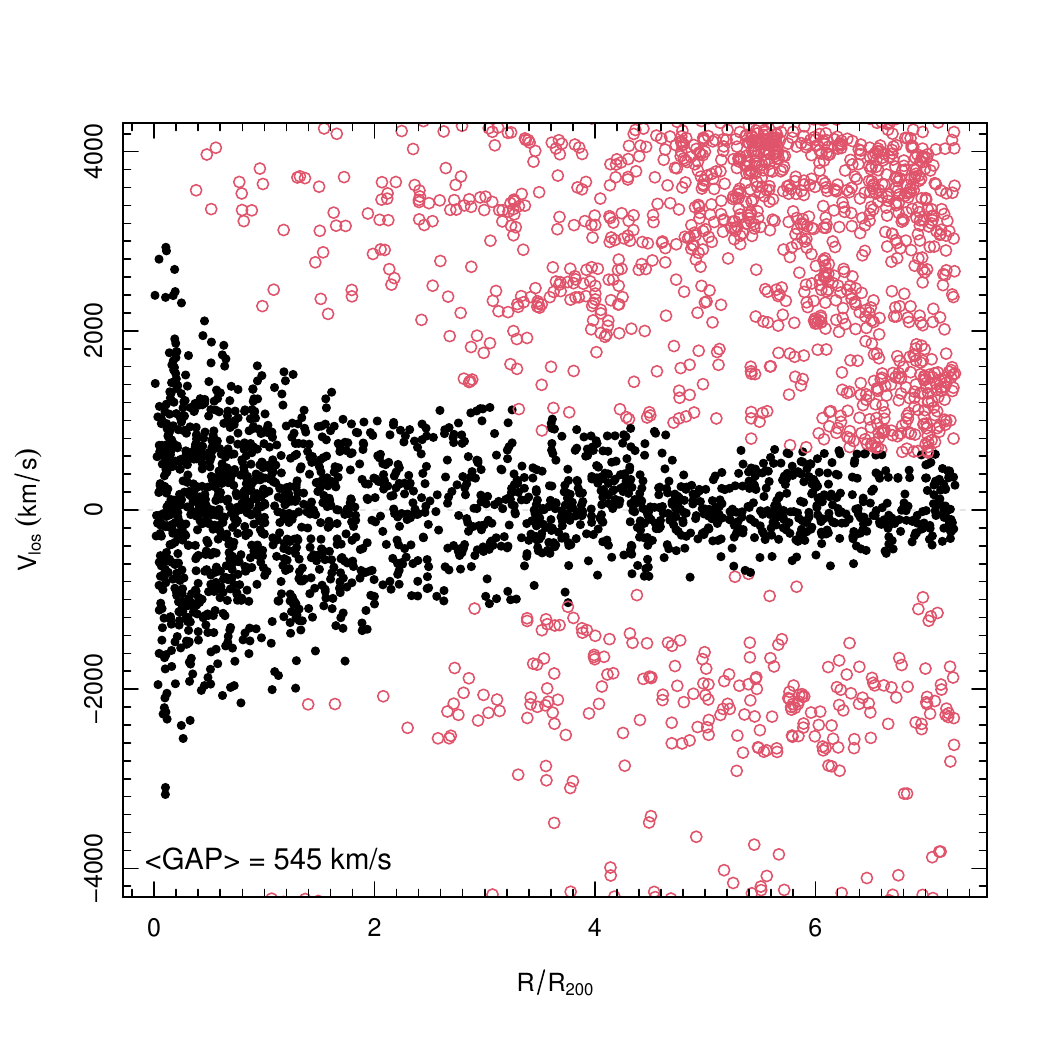}
\vspace{-10pt}
\caption{Projected 
 phase-space of Coma after applying the variable gap method. Member galaxies (1852 objects)
 are presented as filled black circles, and interlopers as open red circles. }
\label{shift}
\end{figure}

The method results in a total of 1852 members and 1456 interlopers in Coma.
Using $R_{200}$ estimated by~\cite{sohn2017velocity} as $R_{200}=2.23_{-0.09}^{+0.08}$ Mpc, we verify that our spatial coverage reaches approximately 8$R_{200}$, which allows for extensive exploration of Coma's projected phase space in the infall regime. See Figure~\ref{shift} for the distribution of members and interlopers in Coma's projected phase space. 
The average gap of $\sim$545 km/s is indicated in the lower-left corner of the figure. This value is close to that used by~\cite{balestra2016clash}. 
\clearpage 

\vspace*{-35pt}

\subsection{Substructure Identification}
\label{subs_sec}

To quantify the impact of low-luminosity galaxies on substructure detection, we adopt an improved version of the Dressler-Schectaman test~\cite{1988AJ.....95..985D}, called DS+~\cite{benavides2023dsp}. Unlike the standard version of the DS test, DS+ not only identifies the presence of substructures in the studied region but also provides the probability (p-value) of each galaxy belonging to the substructure to which it was assigned. After identifying the substructures in each iteration, they are tested by performing 1000 Monte Carlo resampling at the velocity of the galaxies. If the same galaxy is present in more than one substructure, these substructures are excluded to avoid overlap. Additionally, galaxy groups with a high probability of forming a substructure and exhibiting significant proximity in both velocity and distance are merged to prevent fragmentation in the coordinate position space, as long as they do not share any galaxies in common.

In this sense, DS+ checks whether the kinematic parameters of each substructure ($\delta_{v}$ and/or $\delta_{\sigma}$) present significant deviations in relation to the global values of the cluster. To ensure greater precision in the analysis, we only consider substructures with a 99$\%$ probability of belonging to a subgroup (\emph{p}-value $\leq 0.01$), and the host substructure must also have values of $\delta_{v}$ and $\delta_{\sigma}$ with the same level of confidence. Equations~(\ref{ds1}) and (\ref{ds2}) below represent the expressions for $\delta_{v}$ and $\delta_{\sigma}$.


\begin{linenomath}
\begin{equation}
\delta_{v} = N_{g}^{1/2}\, |\,\overline{v_{g}}\,|\,[\,(t_{n}-1)\,\sigma_{v}\,(R_{g})\,]^{-1},
\label{ds1}
\end{equation}
\end{linenomath}


\begin{linenomath}
\begin{equation}
\delta_{\sigma} =  \left[ 1 - \frac{\sigma_{g}}{\sigma_{v}(R)}\right]\,\left\{1 - \left[\frac{(N_{g}-1)}{\chi^{+}_{N_{g}-1} } \right]^{1/2} \right\}^{-1}.
\label{ds2}
\end{equation}
\end{linenomath}

{\noindent
Here, 
 \(R_{g}\) denotes the average projected distance of the substructure from the cluster's center, \(v_{g}\) refers to the mean velocity of the substructure, \(\sigma_{v}(R)\) represents the line-of-sight velocity dispersion profile of the cluster, and \(\sigma_{g}\) corresponds to the line-of-sight velocity dispersion of the substructure. All calculations are performed using galaxy velocities in the cluster's rest frame.}
The Student-\(t\) and \(\chi^2\) distributions are applied to normalize the differences based on the uncertainties in the mean velocity and velocity dispersion, respectively. The parameter \( N_{g} \) specifies the minimum number of galaxies required for a group to be classified as a substructure; otherwise, it indicates the multiplicity of substructures. Here we adopt \( N_{g} =6 \) instead of the ``default'' value \( N_{g} = 3 \). Additional details can be found in~\cite{benavides2023dsp, 2024MNRAS.535.1348C}.

Therefore, the presence of substructures in galaxy clusters reflects dynamically disturbed systems, indicating they are less evolved and non-relaxed, while their absence points to a more stable and advanced evolutionary stage. Thus, identifying substructures is essential, especially when the analysis can extend to deeper samples, incorporating low-flux galaxies.

\subsection{Projected Phase-Space Zones}

To gain valuable insights into the influence of the potentially identified substructures within the cluster, we utilize a machine learning algorithm to reconstruct the orbits of galaxies across various regions of the projected phase-space. The algorithm, named \texttt{ROGER} (Reconstructing Orbits of Galaxies
in Extreme Regions,~\cite{roger}), identifies up to five distinct zones within the projected phase space (PPS) by reconstructing the orbits for each cluster analyzed. The calibrated regions for the algorithm include: (i) Cluster
Ancient Members; (ii) Recent infallers (RIN); (iii)
Backsplash galaxies (BS); (iv) Infalling galaxies (IN); and (v) Interlopers (ITL). Using three distinct machine learning approaches, \texttt{ROGER} identifies galaxy categories in and around clusters based on their PPS coordinates. The algorithm was trained on a galaxy dataset derived from the MDPL2 cosmological simulation and the SAG semi-analytic galaxy formation model~\cite{sag}. For each galaxy, \texttt{ROGER} calculates the likelihood of belonging to one of the five mentioned zones. This classification is vital for investigating how various physical processes shape galaxies and for retracing their past movements within extreme environments like massive galaxy clusters. Among the methods used, K-Nearest Neighbor (KNN) demonstrates the highest accuracy, achieving a sensitivity of 74$\%$ in identifying cluster galaxies categories.

Succinctly, the KNN method uses the Euclidean distance as a metric to measure the separation between two points in a multidimensional space. In the context of galaxy clusters, each galaxy is represented by coordinates in the projected phase-space (PPS), which combines positions and radial velocity information. 

\begin{linenomath}
\begin{equation}
d(\mathbf{x}, \mathbf{y}) = \sqrt{\sum_{i=1}^{n} (x_i - y_i)^2},
\label{eclid}
\end{equation}
\end{linenomath}

\noindent{where,
\( d(\mathbf{x}, \mathbf{y}) \) is the Euclidean distance between the two points \( \mathbf{x} \) and \( \mathbf{y} \),  \( x_i \) and \( y_i \) are the coordinates of the points \( \mathbf{x} \) and \( \mathbf{y} \) in the \(i\)-th dimension, and \( n \) is the number of dimensions, which in the case of galaxy clusters, corresponds to the number of parameters used, such as position and velocity in the PPS.}

The KNN classification algorithm assigns a class to a target point \( \mathbf{x}_{target} \) based on the majority class of its \( k \)-nearest neighbors. The classification rule is given by $\hat{y}_{\text{target}} = \text{mode} \left( y_1, y_2, \dots, y_k \right)$ in which, 
\( \hat{y}_{target} \) is the predicted class label of the target point \( \mathbf{x}_{target} \), \( y_1, y_2, \dots, y_k \) are the class labels of the \( k \)-nearest neighbors to the target point, the function \(\text{mode} \) returns the most frequent class label among the \( k \)-neighbors and the \( k \)-nearest neighbors are determined based on the Euclidean distance between the target point \( \mathbf{x}_{target} \) and all other points in the dataset, selecting the \( k \) closest points.

\section{Results and Analysis}

\begin{figure}[H]
\includegraphics[width=10.5 cm]{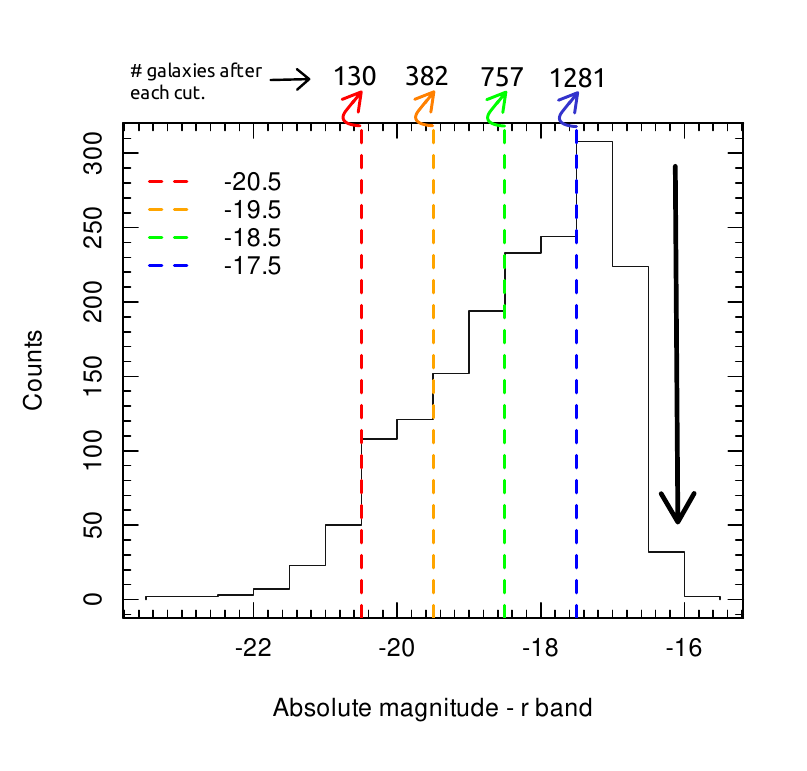}
\caption{Histogram 
 of absolute magnitudes in the r-band with dashed lines indicating the magnitude values used to define the cluster slices. We also show the number of member galaxies after each cut. The vertical downward arrow reflects the detection limit of the instrument for this sample. Below this limit, the probability of identifying low-luminosity galaxies decreases significantly, leading to an incomplete sample, e.g.,~\cite{2023A&A...678A.145M}.}
\label{comafigcuts_ok}
\end{figure}

The purpose of this section is to evaluate the possible dynamical changes of Coma through different magnitude values, going deeper and deeper into the luminosity of the galaxies in order to include low-luminosity galaxies. The chosen magnitude values are: $M_{r} \leq -20.5$, $M_{r} \leq -19.5$, $M_{r} \leq -18.5$ and $M_{r} \leq -17.5$. The following histogram (Figure \ref{comafigcuts_ok}) shows the cutoffs in magnitudes and the threshold from which the magnitude distribution exhibits a sharp drop (vertical down arrow), corresponding to the regime where the statistical contribution of low-luminosity galaxies decreases, due to incompleteness, ensuring the robustness of the sample.
We wish to study the contribution of low-luminosity galaxies to substructure detection.

\subsection{DS+ and Galaxy Distribution}
In each sample, we perform substructure detection using DS+ and then verify their position in the projected phase-space zones determined through the ML \texttt{ROGER} method.
Furthermore, in each panel, the pink dashed line was derived from
$\frac{|V_{\text{los}}|}{\sigma} = -\frac{4}{3} \frac{R}{R_{\text{virial}}} + 2$~(\cite{2013MNRAS.431.2307O}). To account for projection effects in our observational data, we adopted the following assumptions: $R_{\text{virial}} = 2.5 \, R_{200} \quad \text{and} \quad \sigma = \sqrt{3} \, \sigma_{\text{group}}$
Thus, the pink dashed line in the PPS figures serves as a rough indicator to distinguish between recent infallers (\(\tau < 1 \, \text{Gyr}\)) and those who have been infalling for a longer time (\(\tau > 1 \, \text{Gyr}\)), as suggested by~\cite{2017MNRAS.467.4410A}.

\begin{figure}[H]
\begin{minipage}[h]{1\linewidth}
\vspace{-0.3cm}
\begin{minipage}[h]{0.5\linewidth}
\begin{center}
\includegraphics[width=0.95\linewidth]{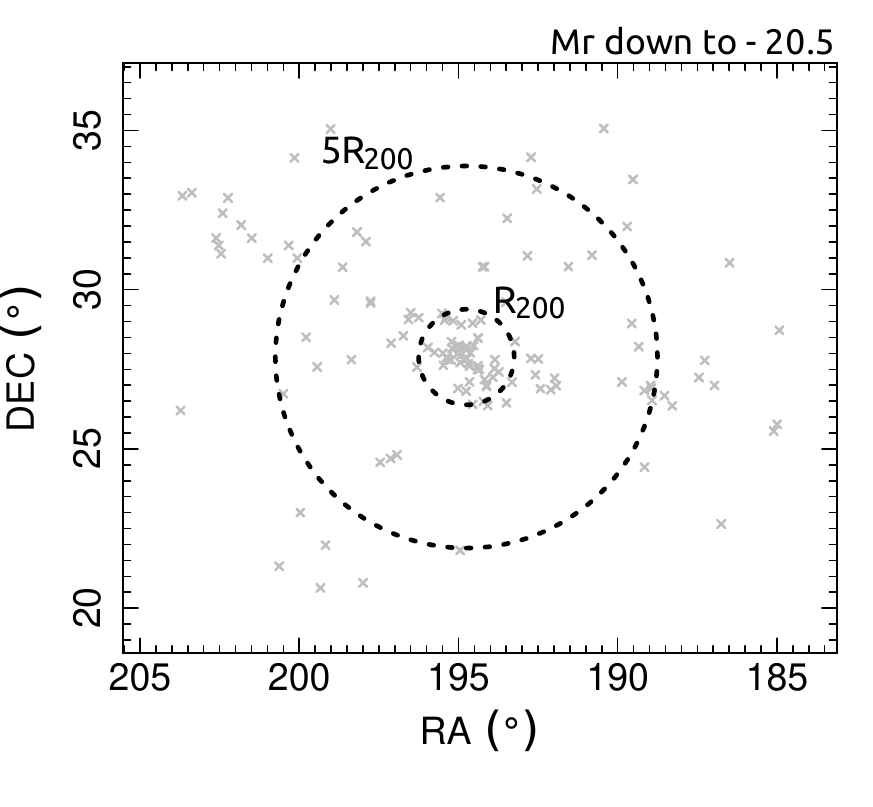} 
\label{qwe1}
\end{center} 
\end{minipage}
\vspace{-0.5cm}
\begin{minipage}[h]{0.5\linewidth}
\begin{center}
\includegraphics[width=0.95\linewidth]{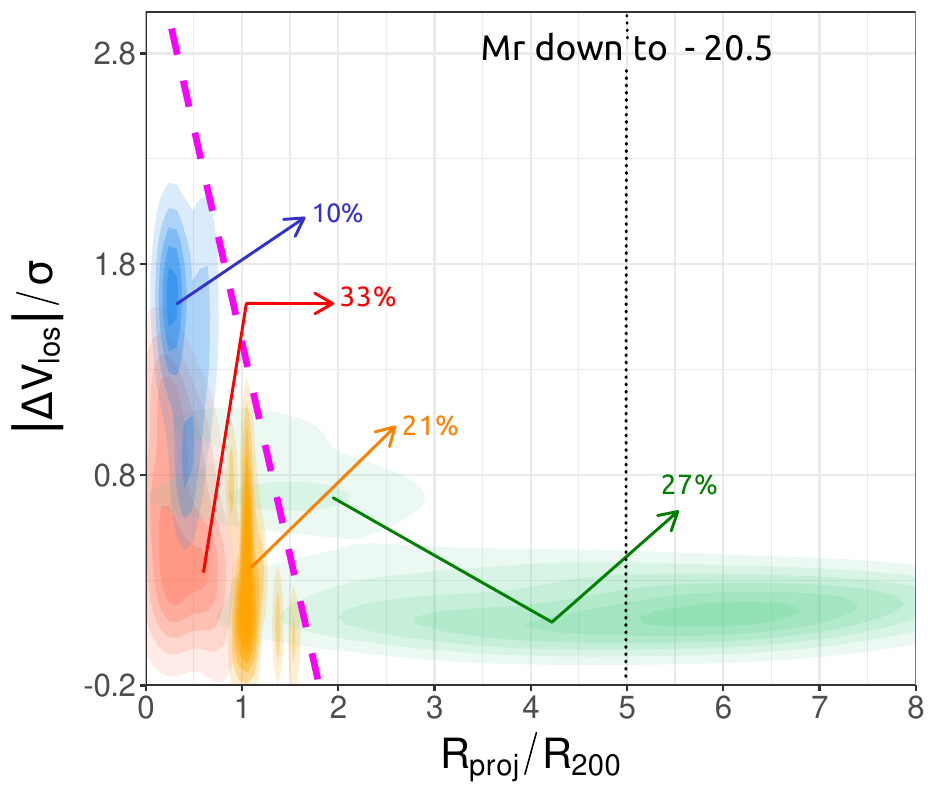} 
\label{qwe1}
\end{center}
\end{minipage}
\vspace{-0.5 cm}
\begin{minipage}[h]{0.5\linewidth}
\begin{center}
\includegraphics[width=.95\linewidth]{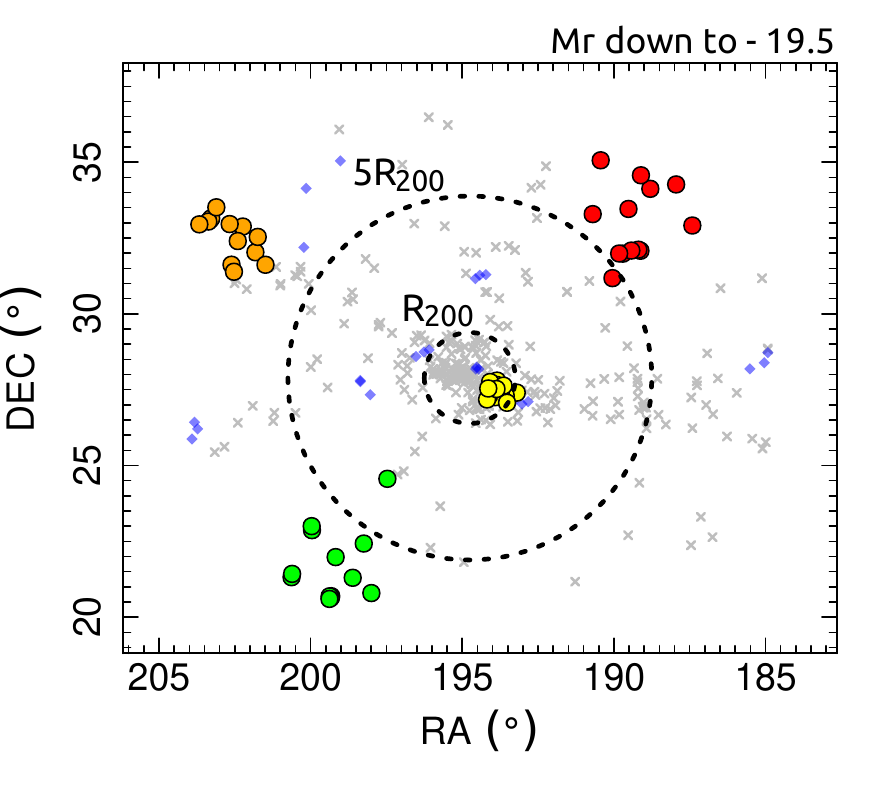} 
\label{qwe1}
\end{center}
\end{minipage}
\begin{minipage}[h]{0.5\linewidth}
\begin{center}
\includegraphics[width=.95\linewidth]{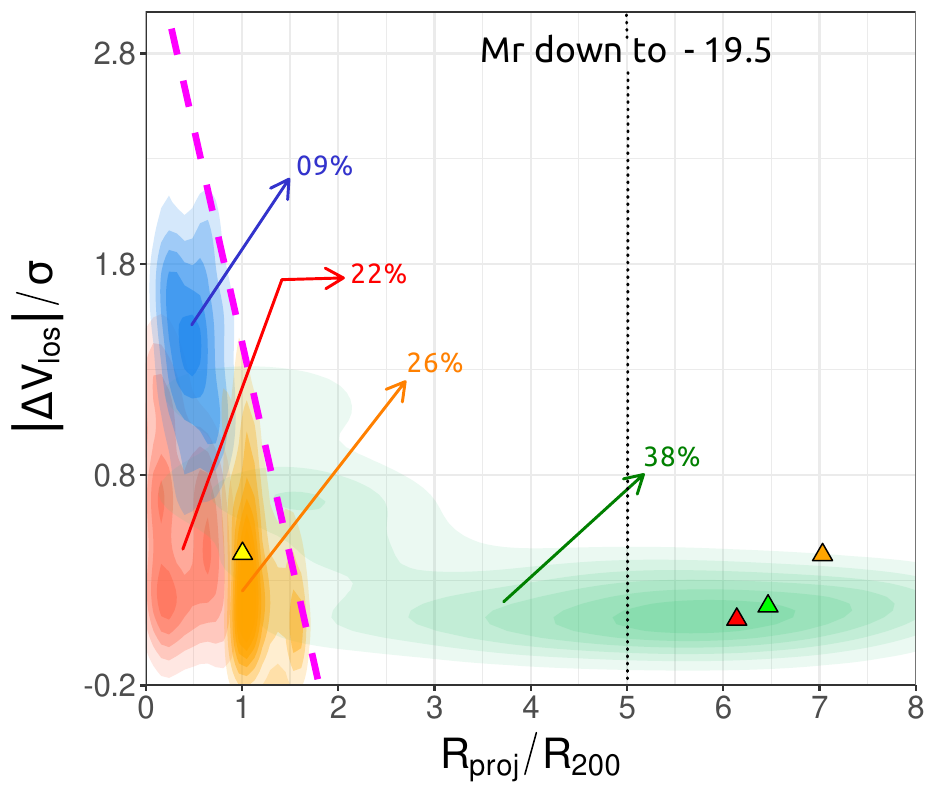} 
\label{qwe1}
\end{center}
\end{minipage}

\vspace{3pt}
\caption{(\textbf{Left 
 column}): The RA-DEC distribution of galaxies in Coma shows field galaxies marked with ``x'' symbols in gray, while statistically significant galaxy groups are represented as colored circles. The size of these circles is proportional to \( 1 - 100 \cdot p \), where \( p \) is the group's \emph{p}-value (see Table~\ref{tab1}). Groups with p-values outside the defined reliability threshold are highlighted as blue diamonds. The dashed circles represent the virial region in Coma \(\sim R_{200} = 2.23 \) Mpc~\cite{sohn2017velocity}, and the turnaround radius, $\sim$5$R_{200}$. (\textbf{Right column}): Projected 
phase-space with regions identified via \texttt{ROGER}. For each region, the percentage of galaxies that make it up in relation to the entire sample is shown. The dashed vertical line indicates the typical value for the turnaround radius, while the triangles over the regions indicate the average position of the identified substructures in the RA-DEC distribution. } 
\label{figsamples1}
\end{minipage}
\end{figure}

Examining the initial results in Figure
~\ref{figsamples1}, which correspond to samples with \mbox{\(M_r \leq -20.5\)} and \(M_r \leq -19.5\), the distribution of objects in the projected phase space highlights spatial and velocity characteristics that align with zones (i) through (iv), as defined in Section~\ref{subs_sec}. These zones are represented by the colors red (i), blue (ii), \mbox{orange (iii),} and green (iv), respectively.
The top panels of Figure~\ref{figsamples1} show that when selecting only the brightest galaxies, with \(M_{r} \leq -20.5\), Coma does not show any evidence of substructures indicated by DS+, although some galaxies are found in subgroups across the infall region of the cluster (green region) according to the work of~\cite{10.1111/j.1365-2966.2011.19756.x}, who also use objects up to $M_r=-20.5$.
However, when considering the cluster with galaxies down to the limit of \( M_r \leq -19.5 \), we identify the presence of four substructures, whose information and properties are detailed in Table~\ref{tab1}. These results are shown in the bottom panels of Figure~\ref{figsamples1}, with the subgroups visualized both in equatorial coordinates and in the PPS. By analyzing the combined projections of the substructures, we observe that three of them are
in the infall zone (bottom right panel) but beyond the turn-around radius, 
$\approx 5R_{200}$ (vertical dotted line in the PPS plot), and hence they are approaching the cluster's global potential for the first time, but they cannot yet be strictly considered as infall subgroups \citep{lopes2024role}; rather, they are systems that are part of the rich large-scale structure around Coma \citep{malavasi2020like}.
At the same time, one substructure (marked in yellow) is located more centrally in the system, within the region delimited by \( R_{200} \) (the internal dashed circle). In the PPS, this substructure is located in the backsplash galaxy region (in orange), in addition to being inside the pink dashed line. According to~\cite{2017MNRAS.467.4410A}, the groups within this line have been experiencing the potential of the cluster for more than 1 Gyr. It is also expected that they have passed at least once through the center of the cluster, e.g.,~\cite{rhee2017phase}.

Another important aspect to consider is the demography of galaxies in the PPS regions. We used a two-sample test of proportions to compare samples with $M_r \leq -20.5$ and $M_r \leq -19.5$. Proportionally, the test indicates more ancient objects in the sample with $M_r \leq -20.5$ and more objects in the first infall zone in the sample with $M_r \leq -19.5$. The other regions of the PPS do not show significant differences in their respective proportions of galaxies at the 95\% confidence level (see Figure~\ref{figsamples1}).

\begin{table}[H] 
\caption{Substructure properties identified by DS+ in Coma up to \(M_{r} = -19.5\). 
\label{tab1}}
\begin{tabularx}{\textwidth}{p{2cm}CCCm{2cm}<{\centering}m{1.5cm}<{\centering}m{1.5cm}<{\centering}}
\toprule
\vspace{-15pt}\textbf{Magnitude Limit} & \textbf{Color Group} & \textbf{N}\boldmath{$_{gal}$} & \textbf{\emph{p}-Value} & \boldmath{$\sigma_{grp}$ } \textbf{[km/s]} & \textbf{RA}\boldmath{$_{grp}$} \textbf{[Degree]} & \textbf{DEC}\boldmath{$_{grp}$} \textbf{[Degree]}\\[-0.5ex]
\midrule
                   & red & 13 & 0.000 & 108.34 $\pm$ 23 & 189.32 & 33.01\\
                   & green & 13 & 0.000 & 105.77 $\pm$ 14 & 199.23 & 21.71\\
M$_r$ $\leq-$19.5 & orange & 14 & 0.001 & 153.95 $\pm$ 08 & 202.63 & 32.55\\
                   & yellow & 09 & 0.007 & 382.96 $\pm$ 57 & 193.82 & 27.44\\
                  
\bottomrule
\end{tabularx}
\noindent{\footnotesize{The columns are: Color group 
 represents the colors of each substructure, only for visual identification, N\(_{gal}\) represents the final number of galaxies in each substructure after validation via Monte Carlo, \emph{p}-value the global significance level of each subgroup considering both deviations (\(\delta_v\) and \(\delta_{\sigma}\)), \(\sigma_{grp}\) the velocity dispersion of the substructures considering the identified member galaxies and RA$_{grp}$ and DEC$_{grp}$ the average positions of the substructures in the sky plane.}}
\end{table}

Next, probing the hierarchical dynamics using galaxies up to \(M_{r} = -18.5\) not only confirms the previous substructures but also includes other groups with equal statistical confidence (top panels of Figure~\ref{figsamples1}). We see in these panels that most of the newly discovered substructures are present in the infall region of the cluster (green region), 
seven of them beyond the turn-around radius, and four that can be considered infall subgroups.
In the central zone of the PPS, we note that
the substructure previously identified in the backsplash region gains new galaxies and is now reclassified as being in the infall region. On the other hand, a new substructure in the backsplash region is found (in purple).
The two-sample proportion test indicates the
demographic similarity between the
samples with $M_r\leq -19.5$ and $M_r \leq -18.5$.
 Table~\ref{tab2} shows the properties of the substructures found \linebreak by DS+. 


\vspace{-6pt}

\begin{table}[H] 
\caption{Substructure properties identified by DS+ in Coma up to \(M_{r} = -18.5\). 
\label{tab2}}
\begin{tabularx}{\textwidth}{p{2cm}m{1.5cm}<{\centering}Cm{1.8cm}<{\centering}m{2cm}<{\centering}m{1.5cm}<{\centering}m{1.5cm}<{\centering}}
\toprule
\vspace{-15pt}\textbf{Magnitude Limit} & \textbf{Color Group} & \textbf{N}\boldmath{$_{gal}$} & \textbf{\emph{p}-Value} & \boldmath{$\sigma_{grp}$ } \textbf{[km/s]} & \textbf{RA}\boldmath{$_{grp}$} \textbf{[Degree]} & \textbf{DEC}\boldmath{$_{grp}$} \textbf{[Degree]}\\[-0.3ex]
\midrule
                   & red & 12 & 0.000 & 117.53 $\pm$ 03 & 190.43 & 34.22\\
                   & green & 10 & 0.000 & 131.29 $\pm$ 11 & 198.77 & 20.56\\
                   & orange & 13 & 0.000 & 223.35 $\pm$ 81 & 202.45 & 32.63\\
                   & yellow & 14 & 0.000 & 334.65 $\pm$ 45 & 198.18 & 28.05\\
                   & cyan & 13 & 0.000 & 283.96 $\pm$ 52 & 188.72 & 25.52\\
                   & magenta & 14 & 0.000 & 154.98 $\pm$ 31 & 201.63 & 26.48\\
M$_r$ $\leq -$18.5 & palegreen & 13 & 0.000 & 418.03 $\pm$ 12 & 195.53 & 32.10\\
                   & gray & 12 & 0.001 & 197.04 $\pm$ 15 & 200.55 & 31.20\\
                   & pink & 13 & 0.001 & 338.83 $\pm$ 95 & 186.05 & 27.17\\
                   & blue & 12 & 0.000 & 157.67 $\pm$ 44 & 188.55 & 31.71\\
                   & purple & 10 & 0.003 & 464.48 $\pm$ 88 & 194.42 & 29.03\\
                   & gold & 23 & 0.007 & 644.16 $\pm$ 51 & 193.08 & 26.72\\
                  
\bottomrule
\end{tabularx}
\noindent{\footnotesize{The columns have the same meanings as in Table~\ref{tab1}.}}

\end{table}

\begin{figure}[H]
\begin{minipage}[h]{1\linewidth}
\vspace{-0.3cm}
\begin{minipage}[h]{0.5\linewidth}
\begin{center}
\includegraphics[width=.95\linewidth]{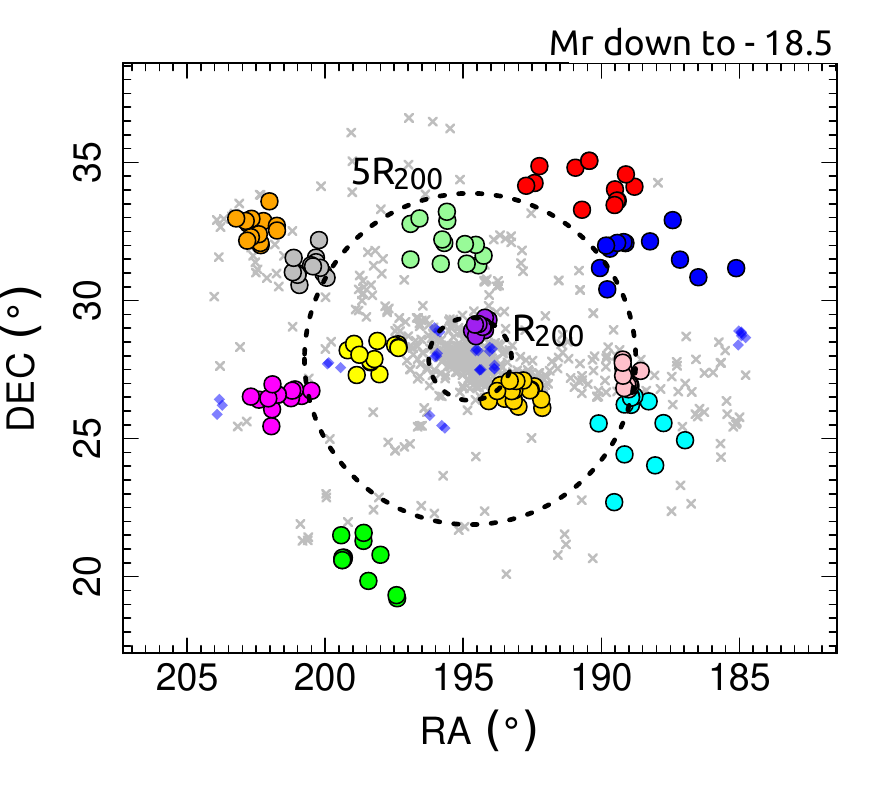} 
\label{qwe1}
\end{center} 
\end{minipage}
\vspace{-0.5cm}
\begin{minipage}[h]{0.5\linewidth}
\begin{center}
\includegraphics[width=.95\linewidth]{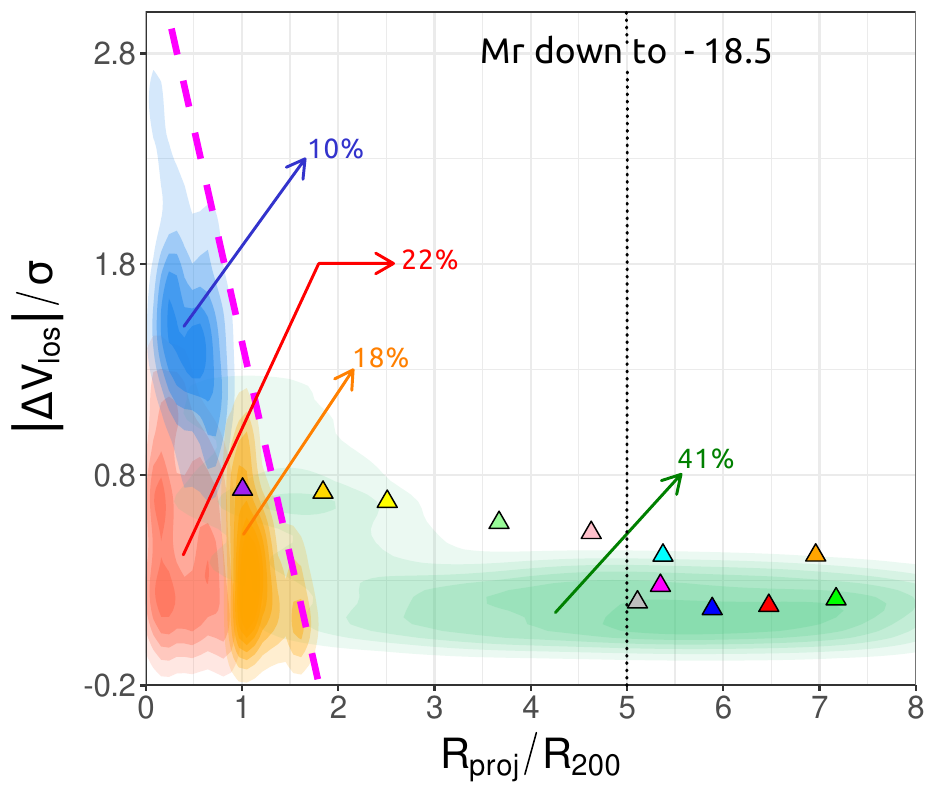} 
\label{qwe1}
\end{center}
\end{minipage}
\vspace{-0.5 cm}
\begin{minipage}[h]{0.5\linewidth}
\begin{center}
\includegraphics[width=.95\linewidth]{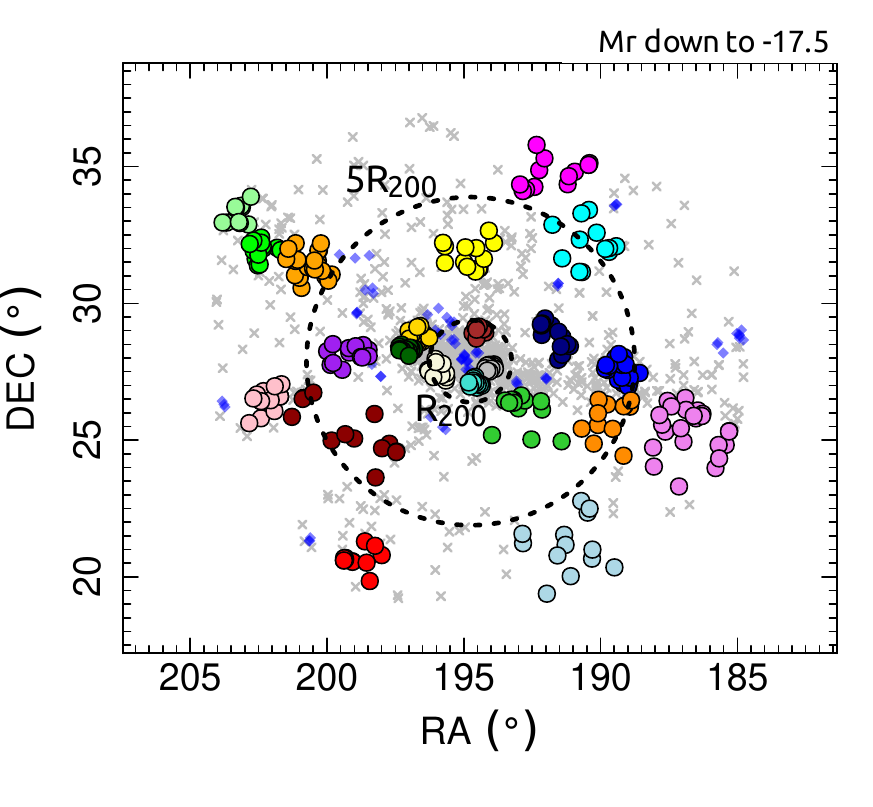} 
\label{qwe1}
\end{center}
\end{minipage}
\begin{minipage}[h]{0.5\linewidth}
\begin{center}
\includegraphics[width=.95\linewidth]{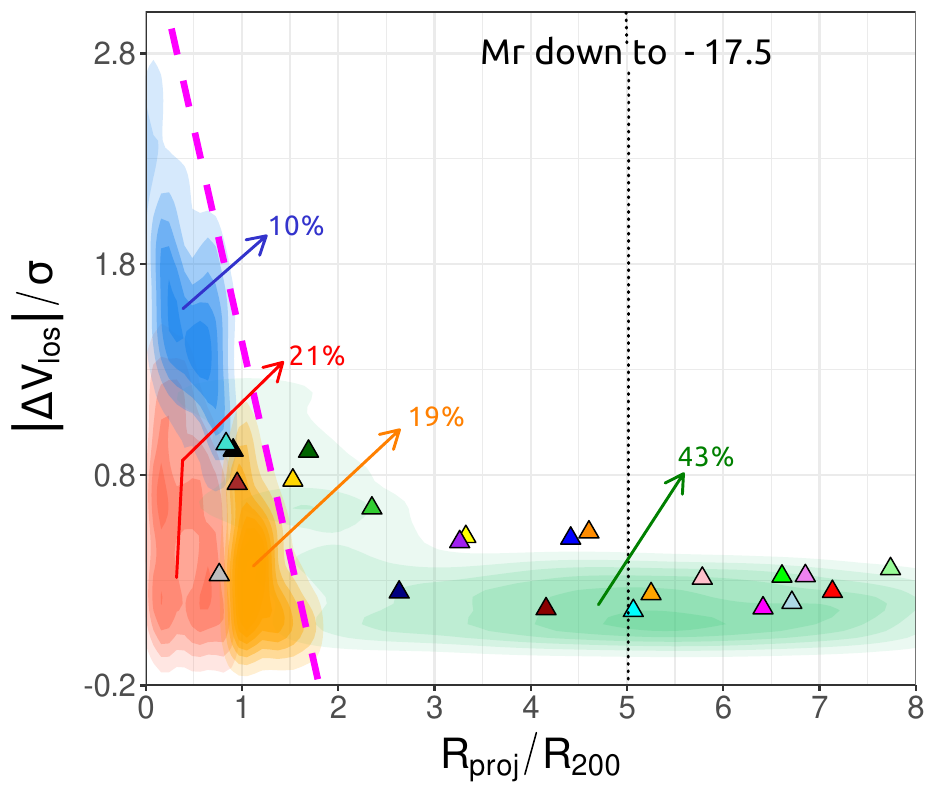} 
\label{qwe1}
\end{center}
\end{minipage}
\vspace{6pt}
\caption{(\textbf{Left 
 column}): The RA-DEC distribution of galaxies in Coma shows field galaxies marked with "x" symbols, while statistically significant galaxy groups are represented as colored circles. (\textbf{Right column}): PPS showing substructures identified at each of the magnitude limits considered, similar to Figure~\ref{figsamples1}.  } 
\label{figsamples2}
\end{minipage}
\end{figure}

Finally, we verify the evolution of the substructures in Coma for galaxies with magnitudes limited to \( M_{r} \leq -17.5 \), corresponding to the faint end of the luminosity function including galaxies generally classified as dwarfs or low luminosities, e.g.,~\cite{2013MNRAS.432.1162L}. 
We note a significant increase in the number of substructures compared to the previous cut, from 12 to 22 in total.
The final count of substructures in Coma is presented in the bottom panels of \linebreak Figure~\ref{figsamples2} and Table~\ref{tab3}. Despite this increase, the demographics in the PPS regions do not change, according to the proportion test at the 95\% confidence level. However, if we compare the first infall region between the first cut ($M_r \leq -20.5$) and the last cut ($M_r \leq -17.5$), we see a significant increase in the occupation of this part of the PPS from 27\% to 43\%, reinforcing how much the periphery of the cluster is more affected by the low-flux population. It is also important to highlight the increase in the number of substructures within the line that separates objects with more than 1 Gyr in the cluster potential. In other words, low-flux galaxies contribute significantly to the dynamic characterization of the cluster, especially because more central substructures are those that actually indicate how regular or not the system is.

Regarding changes in substructures identified in previous magnitude cuts, it is important to note that one of the improvements of DS+ in relation to other versions of the DS algorithm is that the method has important restrictions related to the fragmentation of the structure. As each new magnitude cut adds more galaxies to the sample, it alters the system's overall gravitational potential, thereby impacting the kinematic correlations between the objects. 
This can lead to changes not only in the number of galaxies within the substructures but also in the possibility of merging substructures that are close in projected distance and velocity space into a single entity. Conversely, when new galaxies are added to an already identified substructure, the subgroup's gravitational potential may weaken due to the presence of multiple dominant galaxies, potentially causing the subgroup to split. The condition governing the process described above is expressed by:
\begin{linenomath}
\begin{equation}
d_{i,j} < max(d_{max,i}\,,\, d_{max,j}) \,\, \text{\&}\,\, |\overline{v_{g,i}}-\overline{v_{g,j}}| < max(|v_{max,i}|, |v_{max,j}|),
\label{eclid}
\end{equation}
\end{linenomath}

\noindent{the quantity \( d_{i,j} \) represents the projected distance between the median centers of the groups \( i \) and \( j \), while \( d_{\mathrm{max},i} \) is the maximum distance of any galaxy in the group \( i \) from its group center. Similarly, \( v_{\mathrm{g},i} \) denotes the mean line-of-sight (l.o.s.) velocity of the group \( i \), and \( |v_{\mathrm{max},i}| \) is the maximum absolute velocity difference of any galaxy in group \( i \) relative to its group mean velocity.}

These considerations, along with the p-value criterion - which does not enforce a fixed size for the substructures but requires that the identified groups exhibit significant deviations — can contribute to a degree of similarity in the membership of the group. This is because groups with a large number of members may not be completely homogeneous, leading to their fragmentation into smaller substructures or their exclusion altogether. It is important to highlight that these aspects, previously noted by~\cite{benavides2023dsp,2024MNRAS.535.1348C} and subject to future improvements in DS+ (as discussed in private communication with the author), do not undermine the validity of the detected substructures. The identified substructures remain statistically significant, and the observed dynamical signals are real.



\begin{table}[H] 

\caption{Substructure properties identified by DS+ in Coma up to \(M_{r} = -17.5\). 
\label{tab3}}
 \tablesize{\fontsize{8.5}{8.5}\selectfont}
\begin{tabularx}{\textwidth}{p{1.5cm}m{1.5cm}<{\centering}p{1.2cm}Cm{1.05cm}<{\centering}m{1.75cm}<{\centering}m{1.4cm}<{\centering}m{1.4cm}<{\centering}}
\toprule
\vspace{-11pt}\textbf{Magnitude Limit} & \textbf{Final ID Subs} & \vspace{-11pt}\textbf{Color Group} & \textbf{N}\boldmath{$_{gal}$} & \textbf{\emph{p}-Value} & \boldmath{$\sigma_{grp}$ }\textbf{ [km/s]} & \textbf{RA}\boldmath{$_{grp}$} \textbf{[Degree]} & \textbf{DEC}\boldmath{$_{grp}$} \textbf{[Degree]}\\[-0.3ex]
\midrule
                   & S1 & red & 13 & 0.000 & 145.17 $\pm$ 31 & 198.93 & 20.65\\
                   & S2 & green & 13 & 0.001 & 182.31 $\pm$ 13 & 202.40 & 31.94\\
                   & S3 & orange & 22 & 0.000 & 204.12 $\pm$ 17 & 200.65 & 31.38\\
                   & S4 & yellow & 13 & 0.000 & 310.55 $\pm$ 25 & 194.78 & 31.75\\
                   & S5 & cyan & 12 & 0.000 & 119.36 $\pm$ 22 & 190.38 & 32.18\\
                   & S6 & magenta & 13 & 0.000 & 122.66 $\pm$ 11 & 191.70 & 34.75\\
                   & S7 & palegreen & 11 & 0.000 & 147.16 $\pm$ 21 & 203.26 & 33.23\\
                   & S8 & gray & 13 & 0.000 & 398.76 $\pm$ 15 & 194.03 & 27.65\\
                   & S9 & pink & 12 & 0.000 & 175.47 $\pm$ 15 & 202.20 & 26.45\\
                   & S10 & blue & 23 & 0.000 & 365.30 $\pm$ 04 & 189.26 & 27.53\\
                   & S11 & purple & 13 & 0.000 & 383.66 $\pm$ 17 & 199.17 & 28.13\\
\mbox{\textls[-15]{M$_r$ $\leq -$17.5}} & S12 & gold & 14 & 0.001 & 521.04 $\pm$ 09 & 196.67 & 28.84\\
                   & S13 & brown & 13 & 0.000 & 360.42 $\pm$ 08 & 194.48 & 28.98\\
                   & S14 & darkgreen & 13 & 0.000 & 378.82 $\pm$ 03 & 197.11 & 28.31\\
                   & S15 & lightblue & 12 & 0.000 & 162.03 $\pm$ 04 & 191.12 & 21.17\\
                   & S16 & darkred & 13 & 0.000 & 155.32 $\pm$ 21 & 199.13 & 25.32\\
                   & S17 & darkorange & 12 & 0.000 & 408.40 $\pm$ 18 & 189.59 & 25.83\\
                   & S18 & black & 13 & 0.000 & 412.10 $\pm$ 10 & 195.97 & 27.52\\
                   & S19 & violet & 24 & 0.000 & 296.85 $\pm$ 26 & 186.71 & 25.33\\
                   & S20 & turquoise & 14 & 0.001 & 369.28 $\pm$ 07 & 194.56 & 27.09\\
                   & S21 & limegreen & 13 & 0.001 & 401.93 $\pm$ 17 & 192.87 & 26.08\\
                   & S22 & navy & 12 & 0.007 & 295.04 $\pm$ 19 & 191.67 & 28.80\\
        
\bottomrule
\end{tabularx}
\noindent{\footnotesize{The columns have the same meanings as in Tables~\ref{tab1} and~\ref{tab2}.}}

\end{table}


\subsection{Reliability of Substructures}

An important feature of the method to be mentioned (as verified in~\cite{benavides2023dsp}) is that the galaxies identified in each subclump do not represent the full set of members since DS+ reaches a Completeness (fraction of real members recovered) of around 50\(\%\) as the Richness of the groups varies. Similar Completeness values are also found as a function of \(R/R_{200}\), with a sharp drop from \(1.0\,R/R_{200}\), after reaching \(\sim65\%\) between \(0.8 - 1.0\,R/R_{200}\)---\mbox{Figure 2} 
 in~\cite{benavides2023dsp}. This also may explain the fact that there is no significant increase in the number of galaxies in the substructure for each sample selected in magnitude. This goes against common sense, which would expect an increase in the number of galaxies in substructures as we go deeper into the weak portion of the luminosity function.

\begin{figure}[H]
\begin{minipage}[h]{1\linewidth}
\vspace{-0.3cm}
\begin{minipage}[h]{0.5\linewidth}
\begin{center}
\includegraphics[width=0.98\linewidth]{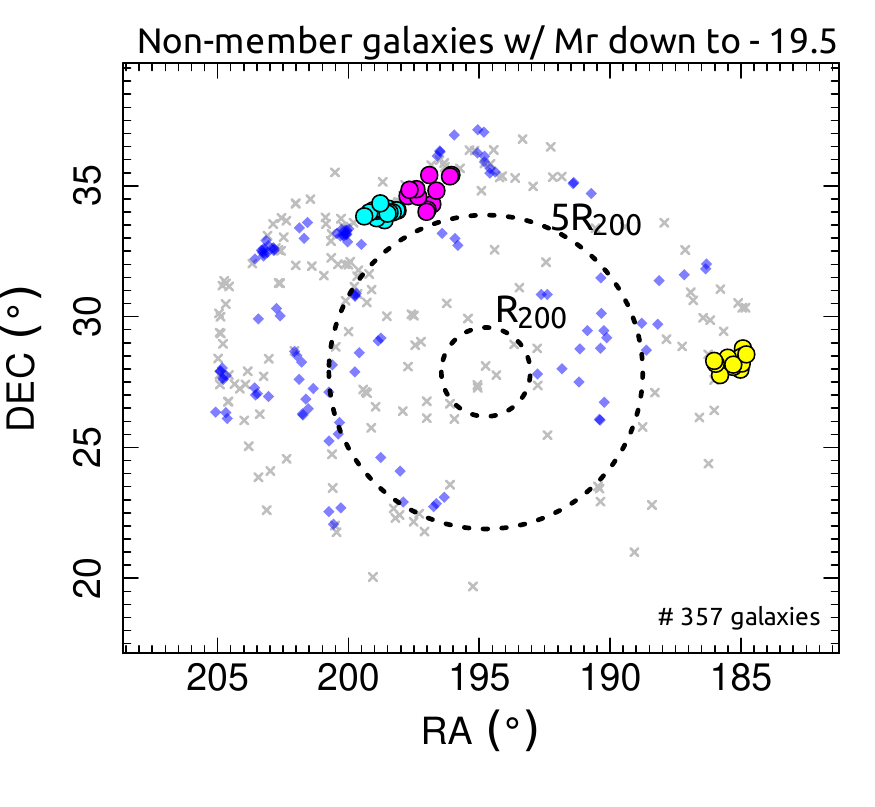} 
\label{qwe1}
\end{center} 
\end{minipage}
\vspace{-0.5cm}
\begin{minipage}[h]{0.5\linewidth}
\begin{center}
\includegraphics[width=.98\linewidth]{figs/mr_19_5.pdf} 
\label{qwe1}
\end{center}
\end{minipage}

\caption{Comparison 
 of the substructure identification regions for samples limited to \(M_{r} \leq - 19.5\). The left panel shows the detection of substructures using DS+ in the sample of 357 non-member galaxies, while the right panel presents the detection in the member galaxy sample, applying the same magnitude limit. 
 The colors between the substructures in both graphs obviously do not represent the same subgroups. The symbols are the same as already explained in Figure \ref{figsamples1}.} 
\label{nomembers}
\end{minipage}
\end{figure}

However, one could argue that such substructures originate from incompleteness in the sky; that is, they are found because they are surrounded by undersampled regions. To solve this possible issue, we study the samples with substructures found in parallel with the corresponding sample of non-members, also detecting subgroups in this sample. The goal of the exercise is to determine whether substructures identified in the member sample region are also present in similar regions of the non-member sample - as long as it has a similar amount of objects---to ensure that the method does not suffer from this effect. For this, we evaluate the first magnitude cut in which DS+ identifies substructures, the sample limited to \(M_{r} \leq -19.5\). Figure~\ref{nomembers} shows the comparison.

As demonstrated in the left panel of Figure~\ref{nomembers}, the DS+ method also identifies three substructures with the same statistical rigor applied to the set of member galaxies (right panel). This comparison reveals that the substructures found in the field of non-member galaxies are located in regions not occupied by any subgroup identified through the magnitude cuts. Additionally, there is a noticeable increase in spurious galaxies (blue diamond-shaped markers), which form groups with lower statistical significance than the established threshold. This outcome is expected, as the sample consists of \linebreak non-member galaxies.


\vspace{3pt}

With these considerations raised previously in mind, it is possible to ensure with some certainty the reality of the identified substructures.


\subsection{Velocity Distributions}

\begin{figure}[H]
\includegraphics[width=13 cm]{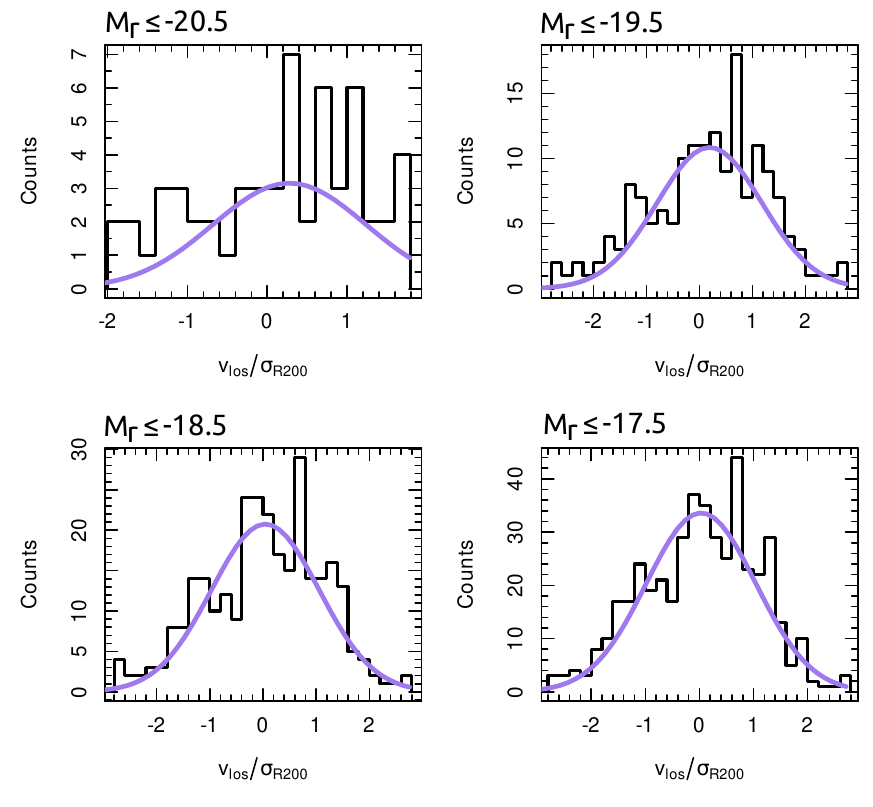}

\vspace{-6pt}
\caption{Frequency 
 histograms with adjusted velocity distribution within the region limited by R200 in the Coma cluster for the different magnitudes analyzed.}
\label{figdistris}
\end{figure}

Additionally, we analyze the velocity distribution of galaxies in the central region of the cluster, particularly within \( R_{200} \), as this type of investigation can reveal crucial information about the system's internal dynamics and its relationship with the presence of faint galaxies e.g.,~\citep{ribeiro2013spider}. This approach becomes even more relevant when considering the impact of galaxy populations with different luminosity limits, 
which potentially bias the analyses. Figure~\ref{figdistris} shows the internal velocity distribution at \(R_{200}\) for each magnitude cut previously considered.

Associated with each velocity distribution, we estimate its kurtosis and skewness since such metrics reveal important insights into the dynamics of clusters, for example~\cite{2024A&A...691A.135B, 2015MNRAS.447.3623V}. Table~\ref{skew_kurt} and Figure~\ref{figdistris} presented highlight the influence of including low-luminosity galaxies in the analysis of velocity distributions, especially within the central region of the cluster. Both the kurtosis (excess kurtosis) and skewness values show significant changes as lower-luminosity galaxies are incorporated into the sample.
With respect to kurtosis, the table shows that all values are negative across all magnitude ranges, indicating distributions that are flatter than a normal distribution. As the magnitude limit increases (\(M_R \leq -20.5\) to \(M_R \leq -17.5\)), kurtosis becomes more negative, reaching a minimum of \(-0.53\) at \(M_R \leq -18.5\). This suggests a greater velocity dispersion, which is expected with the inclusion of low-luminosity galaxies that tend to be associated with substructures~\cite{2012A&A...540A.123E}.


\vspace{-6pt}


\vspace{-6pt}
\begin{table}[H] 
\caption{Skewness and kurtosis for galaxy velocity distributions within \(R_{200}\). 
\label{skew_kurt}}
\begin{tabularx}{\textwidth}{LCC}
\toprule
\textbf{Magnitude Limit} & \textbf{Excess Kurtosis} & \textbf{Skewness} \\
                         & \textbf{\emph{R} \textbf{$\leq$} \emph{R}$_{\textbf{200}}$} & \textbf{\emph{R} \textbf{$\leq$} \emph{R}$_{\textbf{200}}$} \\
\midrule
\(M_{r} \leq -20.5 \) & \(-0.18\) & \(-0.37\) \\
\(M_{r} \leq -19.5 \)& \(-0.36\) & \(-0.39\) \\
\(M_{r} \leq -18.5 \)& \(-0.53\) & \(-0.29\) \\
\(M_{r} \leq -17.5 \)& \(-0.46\) & \(-0.24\) \\
\bottomrule
\end{tabularx}
\noindent{\footnotesize{The columns are: Excess Kurtosis
 measures the tail behavior of the velocity distribution relative to a normal distribution, and Skewness indicates the asymmetry of the velocity distribution.}}

\end{table}


For skewness, all values are also negative, indicating a slight leftward asymmetry in the velocity distribution. This supports not only the presence of substructures but also the peculiar flow of galaxies from the cluster's outskirts, e.g., \cite{2017AJ....154...96D}. This result is further corroborated through the analysis of the projected phase space (PPS), where a significant amount of substructures is observed in the periphery of the systems, inside and outside the turn-around radius. The analysis emphasizes the crucial importance of including low-luminosity galaxies in dynamical studies of galaxy clusters, as they have the ability to reveal complex interactions that are often obscured in analyses focusing solely on brighter magnitude limits, thereby excluding the faint portions of the luminosity function.


\section{Discussion}

This study emphasizes the critical role of low-luminosity galaxy populations in shaping the dynamics of galaxy clusters. Using a robust methodology for the analysis of dynamic indicators, combined with carefully selected sub-samples of the Coma cluster, we initially find that limiting the sample to the commonly used magnitude threshold in SDSS cluster studies at $z < 0.1$, $M_r \leq -20.5$, 
Coma does not exhibit any significant substructures. Specifically, there are no kinematical deviations from the global field, nor are there substantial velocity dispersion changes relative to the entire system.

However, as we extend our analysis to include less luminous galaxies, initially with a threshold of $M_r = -19.5$, we detect the first signs of an association between this population and the presence of substructures. We identify four subgroups: one located within $R_{200}$, situated in the backsplash region of the phase-space diagram (PPS), and three others in the first infall zone of the PPS but beyond the $5R_{200}$ boundary. The kurtosis and skewness values further support a statistically suggestive (albeit limited) relationship between the inclusion of low-luminosity galaxies and the cluster's peripheral dynamics. For instance, galaxies moving towards the cluster center---characteristic of less luminous galaxies in peripheral or halo regions---contribute to the observed negative skewness.

For galaxies with magnitudes up to \(-18.5\) and \(-17.5\), our results are clear and compelling. We observe a significant number of substructures, not only in the first infall region (highlighted by the green zone in the PPS) but also within the central region of the cluster, bounded by $R_{200}$. In this central region, we identify four substructures, each with notable dynamical consequences for the overall system. This highlights an important point: less luminous galaxies are capable of tracing not only the peripheral areas of the clusters but also regions within the halo, where the dynamical effects of substructures may be less pronounced. These effects are often mitigated by the strong gravitational interactions with the dominant cluster potential. Thus, the inclusion of low-luminosity galaxies in cluster analyses not only refines estimates of the cluster's mass and substructure but also enhances our understanding of the dynamic processes shaping these colossal systems. This more complex scenario for Coma dynamics is in agreement with studies performed in X-rays \citep{2021A&A...651A..41C}. Furthermore, low-flux objects allow for better mapping of subgroups in the infall regime around the cluster, allowing a better understanding of the preprocessing process that may be underway around Coma. Finally, we note that demographics in the PPS regions of the cluster stabilize approximately at a cutoff $M_r \leq -19.5$.

Thus, it becomes evident that low-luminosity galaxies play a fundamental role in understanding the dynamical state of galaxy clusters. These galaxies, often overlooked because of their faint nature, serve as sensitive tracers of substructure dynamics and provide crucial insights into the cluster's evolutionary history. Possibly, low-luminosity galaxies can reveal previously undetected substructures because they are more susceptible to interactions within the cluster environment, as our results suggest. Indeed, we show that the absence of low-flux objects can lead to a significant underestimate of the number of substructures in clusters, both in their most central parts, in the infall region, and beyond, connecting to the large-scale structure. The difference in results for cuts at $M_r \leq -20.5$ and $M_r \leq -17.5$ leads to dramatically different results, indicating either no substructure or \linebreak 22 subgroups distributed across vast portions of the PPS of the Coma cluster.

\begin{figure}[H]

\begin{adjustwidth}{-\extralength}{0cm}
\centering 
\includegraphics[width=15cm, angle = 0]{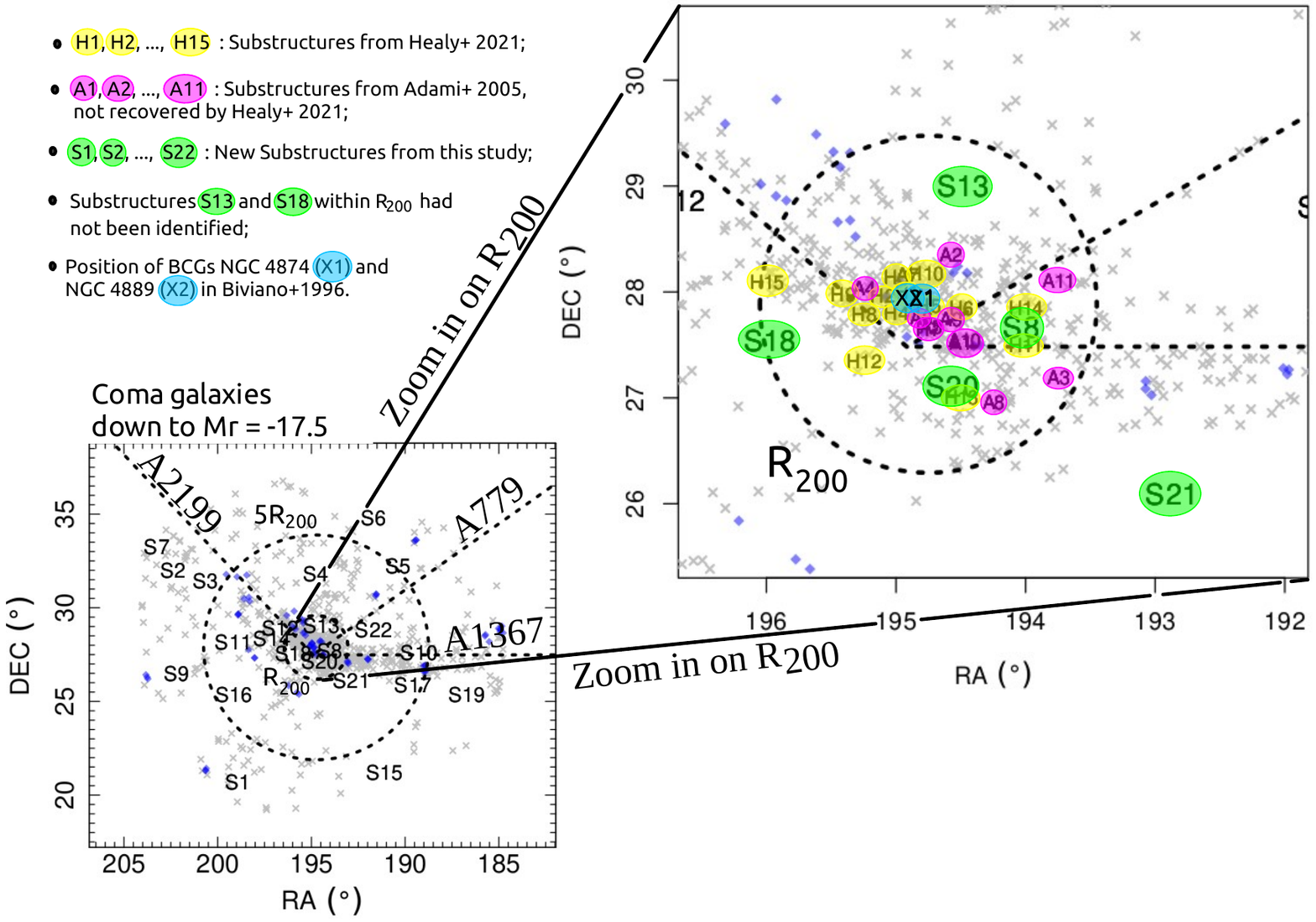}
\end{adjustwidth}
\caption{Zoom 
 in of the region bounded by \(R_{200}\) in the RA $\times$ DEC plane for the Coma cluster used in this study. The acronyms A1 to A11 (in pink) and H1 to H15 (in yellow) represent the substructures identified by~\cite{2005A&A...443...17A,2021A&A...650A..76H}. The groups found in this work are represented by S1 to S22 in green. The dashed lines aligned with the center of Coma show the direction of three clusters in its vicinity. X1 and X2 in light blue represent the positions of the BCGs in the~\cite{1996A&A...311...95B} study.}
\label{comasuper}
\end{figure}

Simultaneously, we compare our study with similar works conducted on the Coma cluster, specifically~\cite{2005A&A...443...17A} (Adami and 2005) and~\cite{2021A&A...650A..76H} (Healy and 2021). Although both studies identify substructures within the region limited to \( R_{200} \) of Coma, they employ different methodologies. Where~\cite{2005A&A...443...17A}\endnote{The authors selected galaxies with \( R < 13 \) and a color index \( B - R = 1.5 \). After applying the necessary approximations to convert between photometric systems, this corresponds to an absolute magnitude of \( M_r \sim -22.06 \).} use 
 a hierarchical approach based on the relative binding energy of galaxies (SG method), ~\cite{2021A&A...650A..76H} apply the classical DS test to a set of redshifts predominantly from SDSS DR13, considering only galaxies with \( r > 17.7 \), which corresponds to an approximate absolute magnitude of \( M_r \sim -17.3 \). Figure~\ref{comasuper} illustrates the comparison.
We first present the final distribution of the 22 substructures in the \( \text{RA} \times \text{DEC} \) plane of Coma for the last magnitude cut (\( M_r \leq -17.5 \)). The substructures are shown by their mean positions and labeled with their respective IDs, ranging from S1 to S22, as detailed in Table~\ref{tab3}. Next, we zoom in on the central region to superimpose our substructures onto those identified by~\cite{2005A&A...443...17A} (A1 to A11) and~\cite{2021A&A...650A..76H} (H1 to H15) in their studies. To avoid visual clutter, we display only the substructures identified by~\cite{2005A&A...443...17A} that were not recovered by~\cite{2021A&A...650A..76H}. Additionally, we include dashed lines extending from the center of Coma, representing the directions to nearby clusters connected to Coma. We should also note that the central substructures S13 and S18 have probably been detected for the first time in this work since their locations do not overlap with those previously reported in the literature, while the substructures S8 and S20 cannot be considered new, as they occupy regions close to other substructures already identified by~\cite{2021A&A...650A..76H} (S8 near H11 and H14; S20 near H13).
These findings highlight the significance of our methodological robustness, which not only employed a strict magnitude threshold \(M_{r} \leq -17.5\) but also achieved a higher sensitivity in identifying previously undetected (and possibly already detected) structures. 

Finally, we also investigated whether the results found by~\cite{1996A&A...311...95B} (Biviano + 1996) are reproduced in our substructures: the fact that its bright galaxies (\(M_{B} \leq 17 \)) tend to be grouped in substructures or around dominant galaxies such as NGC 4874 and NGC 4889---indicating that these galaxies tend to group around the most massive galaxies in the cluster---while the fainter ones (\( 17 < M_{B} \leq 20 \)) have a more uniform distribution (suggesting that they better trace the general distribution of matter in the cluster, forming more continuous structures). 
To do this, once we have the absolute magnitudes in the r-band for our sample of galaxies, we used the characteristic absolute magnitude obtained in the B-band (\(M_{B}^{*} \approx -19.8 \)) for a sample of $\sim 1628$ galaxies in Coma (\cite{2002MNRAS.329..385B}) and then we obtained the corresponding characteristic magnitude in the r-band with value of \(M_{r}^{*}\approx -18.6 \), using 1.2 as the approximate difference between the B and r bands (\cite{1995PASP..107..945F}).

Therefore, in our sample of galaxies down to -17.5 was defined as bright, objects with \(M_{r} \leq -18.6 \) (697 galaxies) and faint, those with \(-18.6 < M_{r} \leq -17.5\) (584 galaxies). In the end, we observe that of the 311 galaxies associated with substructures up to \( M_{r} \leq -17.5 \), only \(\approx\)45\% are galaxies belonging to the bright sample, while \(65\%\) are of the faint type. For the 426 galaxies within \(R_{200}\), \(62\%\) are galaxies considered bright.
Overall, while these findings should be interpreted with caution, they do not necessarily contradict the results of Biviano+1996. The differences in methodology for identifying substructures and sample selection criteria, along with the fact that Biviano+1996 focused on a region limited to \linebreak 1500 arcsec around the center of Coma (\( \approx\)1.0 Mpc h
$^{-1}$), make direct comparisons challenging. However, these preliminary results suggest that the substructures may be more evolved, having already incorporated a larger fraction of faint galaxies. This indicates a different stage in the cluster's evolution, even though the tendency of bright galaxies to be located near dominant galaxies within \(R_{200}\) remains consistent. In Figure~\ref{comasuper}, we also show the positions of the BCGs around which Biviano+1996 identified a larger grouping of galaxies considered bright.


\authorcontributions{{The authors contributed equally to this work.}}

\funding{{~~}}

\dataavailability{We used public data from SDSS (Sloan Digital Sky Survey, \url{https://www.sdss.org} (accessed on 28 November 2024))
.} 

\acknowledgments{{We} 
 are grateful to the reviewers for their thorough evaluation of our manuscript and their
valuable suggestions, which have helped improve our work.
The authors also thank Silas S. Santos for important contributions to the shifting-gapper method.
FRMN thanks the financial support of the Coordenaço de Aperfeiçoamento de Pessoal de Nvel Superior---Brasil (CAPES)---Finance Code 001. 
ALBR thanks CNPq for their support, grant 316317/2021-7 and FAPESB INFRA PIE 0013/2016.}

\conflictsofinterest{ The authors declare no conflicts of interest.} 

\begin{adjustwidth}{-\extralength}{0cm}
\printendnotes[custom] 

\reftitle{References}


\bibliography{universe-3469021}

\begin{thebibliography}{999}

\bibitem[{De Lucia} et~al.(2004){De Lucia}, {Kauffmann}, {Springel}, {White}, {Lanzoni}, {Stoehr}, {Tormen}, and {Yoshida}]{2004MNRAS.348..333D}
{De Lucia}, G.; {Kauffmann}, G.; {Springel}, V.; {White}, S.D.M.; {Lanzoni}, B.; {Stoehr}, F.; {Tormen}, G.; {Yoshida}, N.
\newblock {Substructures in cold dark matter haloes}.
\newblock {\em Monthly Notices of the Royal Astronomical Society} {\bf 2004}, {\em 348},~333--344.
\newblock {\url{https://doi.org/10.1111/j.1365-2966.2004.07372.x}}.

\bibitem[{Old} et~al.(2018){Old}, {Wojtak}, {Pearce}, {Gray}, {Mamon}, {Sif{\'o}n}, {Tempel}, {Biviano}, {Yee}, {de Carvalho}, {M{\"u}ller}, {Sepp}, {Skibba}, {Croton}, {Bamford}, {Power}, {von der Linden}, and {Saro}]{2018MNRAS.475..853O}
{Old}, L.; {Wojtak}, R.; {Pearce}, F.R.; {Gray}, M.E.; {Mamon}, G.A.; {Sif{\'o}n}, C.; {Tempel}, E.; {Biviano}, A.; {Yee}, H.K.C.; {de Carvalho}, R.;  et~al.
\newblock {Galaxy Cluster Mass Reconstruction Project - III. The impact of dynamical substructure on cluster mass estimates}.
\newblock {\em Monthly Notices of the Royal Astronomical Society} {\bf 2018}, {\em 475},~853--866.
\newblock {\url{https://doi.org/10.1093/mnras/stx3241}}.

\bibitem[{Whiley} et~al.(2008){Whiley}, {Arag{\'o}n-Salamanca}, {De Lucia}, {von der Linden}, {Bamford}, {Best}, {Bremer}, {Jablonka}, {Johnson}, {Milvang-Jensen}, {Noll}, {Poggianti}, {Rudnick}, {Saglia}, {White}, and {Zaritsky}]{2008MNRAS.387.1253W}
{Whiley}, I.M.; {Arag{\'o}n-Salamanca}, A.; {De Lucia}, G.; {von der Linden}, A.; {Bamford}, S.P.; {Best}, P.; {Bremer}, M.N.; {Jablonka}, P.; {Johnson}, O.; {Milvang-Jensen}, B.;  et~al.
\newblock {The evolution of the brightest cluster galaxies since z \raisebox{-0.5ex}\textasciitilde 1 from the ESO Distant Cluster Survey (EDisCS)}.
\newblock {\em Monthly Notices of the Royal Astronomical Society} {\bf 2008}, {\em 387},~1253--1263.
\newblock {\url{https://doi.org/10.1111/j.1365-2966.2008.13324.x}}.

\bibitem[{Shen} et~al.(2014){Shen}, {Yang}, {Mo}, {van den Bosch}, and {More}]{2014ApJ...782...23S}
{Shen}, S.; {Yang}, X.; {Mo}, H.; {van den Bosch}, F.; {More}, S.
\newblock {The Statistical Nature of the Brightest Group Galaxies}.
\newblock {\em The Astrophysical Journal} {\bf 2014}, {\em 782},~23.
\newblock {\url{https://doi.org/10.1088/0004-637X/782/1/23}}.

\bibitem[Mason et~al.(2023)Mason, Trenti, and Treu]{mason2023brightest}
Mason, C.A.; Trenti, M.; Treu, T.
\newblock The brightest galaxies at cosmic dawn.
\newblock {\em Monthly Notices of the Royal Astronomical Society} {\bf 2023}, {\em 521},~497--503.

\bibitem[Kaviraj et~al.(2011)Kaviraj, Tan, Ellis, and Silk]{kaviraj2011coincidence}
Kaviraj, S.; Tan, K.M.; Ellis, R.S.; Silk, J.
\newblock A coincidence of disturbed morphology and blue UV colour: minor-merger-driven star formation in early-type galaxies at z $\sim$ 0.6.
\newblock {\em Monthly Notices of the Royal Astronomical Society} {\bf 2011}, {\em 411},~2148--2160.

\bibitem[{Costa} et~al.(2018){Costa}, {Ribeiro}, and {de Carvalho}]{2018MNRAS.473L..31C}
{Costa}, A.P.; {Ribeiro}, A.L.B.; {de Carvalho}, R.R.
\newblock {The shape of velocity dispersion profiles and the dynamical state of galaxy clusters}.
\newblock {\em Monthly Notices of the Royal Astronomical Society} {\bf 2018}, {\em 473},~L31--L35.
\newblock {\url{https://doi.org/10.1093/mnrasl/slx156}}.

\bibitem[Tanaka et~al.(2004)Tanaka, Goto, Okamura, Shimasaku, and Brinkmann]{tanaka2004environmental}
Tanaka, M.; Goto, T.; Okamura, S.; Shimasaku, K.; Brinkmann, J.
\newblock The environmental dependence of galaxy properties in the local universe: dependences on luminosity, local density, and system richness.
\newblock {\em The Astronomical Journal} {\bf 2004}, {\em 128},~2677.

\bibitem[{Berg} et~al.(2012){Berg}, {Skillman}, {Marble}, {van Zee}, {Engelbracht}, {Lee}, {Kennicutt}, {Calzetti}, {Dale}, and {Johnson}]{2012ApJ...754...98B}
{Berg}, D.A.; {Skillman}, E.D.; {Marble}, A.R.; {van Zee}, L.; {Engelbracht}, C.W.; {Lee}, J.C.; {Kennicutt}, Robert~C., J.; {Calzetti}, D.; {Dale}, D.A.; {Johnson}, B.D.
\newblock {Direct Oxygen Abundances for Low-luminosity LVL Galaxies}.
\newblock {\em Astrophysical Journal} {\bf 2012}, {\em 754},~98.
\newblock {\url{https://doi.org/10.1088/0004-637X/754/2/98}}.

\bibitem[{Fitchett} and {Webster}(1987)]{1987ApJ...317..653F}
{Fitchett}, M.; {Webster}, R.
\newblock {Substructure in the Coma Cluster}.
\newblock {\em Astrophysical Journal} {\bf 1987}, {\em 317},~653.
\newblock {\url{https://doi.org/10.1086/165310}}.

\bibitem[{Gambera} et~al.(1997){Gambera}, {Pagliaro}, {Antonuccio-Delogu}, and {Becciani}]{1997ApJ...488..136G}
{Gambera}, M.; {Pagliaro}, A.; {Antonuccio-Delogu}, V.; {Becciani}, U.
\newblock {A Three-dimensional Wavelet Analysis of Substructure in the Coma Cluster: Statistics and Morphology}.
\newblock {\em The Astrophysical Journal} {\bf 1997}, {\em 488},~136--145.
\newblock {\url{https://doi.org/10.1086/304684}}.

\bibitem[{Adami} et~al.(2005){Adami}, {Biviano}, {Durret}, and {Mazure}]{2005A&A...443...17A}
{Adami}, C.; {Biviano}, A.; {Durret}, F.; {Mazure}, A.
\newblock {The build-up of the Coma cluster by infalling substructures}.
\newblock {\em Astronomy \& Astrophysics} {\bf 2005}, {\em 443},~17--27.
\newblock {\url{https://doi.org/10.1051/0004-6361:20053504}}.

\bibitem[{Churazov} et~al.(2021){Churazov}, {Khabibullin}, {Lyskova}, {Sunyaev}, and {Bykov}]{2021A&A...651A..41C}
{Churazov}, E.; {Khabibullin}, I.; {Lyskova}, N.; {Sunyaev}, R.; {Bykov}, A.M.
\newblock {Tempestuous life beyond R$_{500}$: X-ray view on the Coma cluster with SRG/eROSITA. I. X-ray morphology, recent merger, and radio halo connection}.
\newblock {\em Astronomy \& Astrophysics} {\bf 2021}, {\em 651},~A41.
\newblock {\url{https://doi.org/10.1051/0004-6361/202040197}}.

\bibitem[Fadda et~al.(1996)Fadda, Girardi, Giuricin, Mardirossian, and Mezzetti]{Fadda1996}
Fadda, D.; Girardi, M.; Giuricin, G.; Mardirossian, F.; Mezzetti, M.
\newblock The dynamics of galaxy clusters.
\newblock {\em The Astrophysical Journal} {\bf 1996}, {\em 473},~670--685.
\newblock {\url{https://doi.org/10.1086/178181}}.

\bibitem[Girardi et~al.(1996)Girardi, Fadda, Giuricin, Mardirossian, Mezzetti, and Biviano]{girardi1996velocity}
Girardi, M.; Fadda, D.; Giuricin, G.; Mardirossian, F.; Mezzetti, M.; Biviano, A.
\newblock Velocity Dispersions and X-Ray Temperatures of Galaxy Clusters.
\newblock {\em Astrophysical Journal v. 457, p. 61} {\bf 1996}, {\em 457},~61.

\bibitem[Lopes et~al.(2009)Lopes, De~Carvalho, Kohl-Moreira, and Jones]{lopes2009nosocs}
Lopes, P.; De~Carvalho, R.; Kohl-Moreira, J.; Jones, C.
\newblock NoSOCS in SDSS--I. Sample definition and comparison of mass estimates.
\newblock {\em Monthly Notices of the Royal Astronomical Society} {\bf 2009}, {\em 392},~135--152.

\bibitem[Adami et~al.(1998)Adami, Mazure, Biviano, Katgert, and Rhee]{adami1998eso}
Adami, C.; Mazure, A.; Biviano, A.; Katgert, P.; Rhee, G.
\newblock The ESO nearby Abell cluster survey. IV. The fundamental plane of clusters of galaxies.
\newblock {\em Astronomy and Astrophysics, v. 331, p. 493-505 (1998)} {\bf 1998}, {\em 331},~493--505.

\bibitem[Sohn et~al.(2017)Sohn, Geller, Zahid, Fabricant, Diaferio, and Rines]{sohn2017velocity}
Sohn, J.; Geller, M.J.; Zahid, H.J.; Fabricant, D.G.; Diaferio, A.; Rines, K.J.
\newblock The velocity dispersion function of very massive galaxy clusters: Abell 2029 and Coma.
\newblock {\em The Astrophysical Journal Supplement Series} {\bf 2017}, {\em 229},~20.

\bibitem[Balestra et~al.(2016)Balestra, Mercurio, Sartoris, Girardi, Grillo, Nonino, Rosati, Biviano, Ettori, Forman, et~al.]{balestra2016clash}
Balestra, I.; Mercurio, A.; Sartoris, B.; Girardi, M.; Grillo, C.; Nonino, M.; Rosati, P.; Biviano, A.; Ettori, S.; Forman, W.;  et~al.
\newblock CLASH-VLT: Dissecting the Frontier Fields Galaxy Cluster MACS J0416. 1-2403 with~ 800 Spectra of Member Galaxies.
\newblock {\em The Astrophysical Journal Supplement Series} {\bf 2016}, {\em 224},~33.

\bibitem[{Dressler} and {Shectman}(1988)]{1988AJ.....95..985D}
{Dressler}, A.; {Shectman}, S.A.
\newblock {Evidence for substructure in rich clusters of galaxies from radial-velocity measurements}.
\newblock {\em Astronomical Journal} {\bf 1988}, {\em 95},~985--995.
\newblock {\url{https://doi.org/10.1086/114694}}.

\bibitem[{Benavides} et~al.(2023){Benavides}, {Biviano}, and {Abadi}]{benavides2023dsp}
{Benavides}, J.A.; {Biviano}, A.; {Abadi}, M.G.
\newblock {DS+: A method for the identification of cluster substructures}.
\newblock {\em Astronomy \& Astophysics} {\bf 2023}, {\em 669},~A147.
\newblock {\url{https://doi.org/10.1051/0004-6361/202245422}}.

\bibitem[{Costa} et~al.(2024){Costa}, {Ribeiro}, {de Carvalho}, and {Benavides}]{2024MNRAS.535.1348C}
{Costa}, A.P.; {Ribeiro}, A.L.B.; {de Carvalho}, R.R.; {Benavides}, J.A.
\newblock {Boosting the evolutionary picture of Cl 0024+17 and MS 0451-03: a case study at intermediate-redshift}.
\newblock {\em Monthly Notices of the Royal Astronomical Society} {\bf 2024}, {\em 535},~1348--1363.
\newblock {\url{https://doi.org/10.1093/mnras/stae2410}}.

\bibitem[{de los Rios} et~al.(2021){de los Rios}, {Mart{\'\i}nez}, {Coenda}, {Muriel}, {Ruiz}, {Vega-Mart{\'\i}nez}, and {Cora}]{roger}
{de los Rios}, M.; {Mart{\'\i}nez}, H.J.; {Coenda}, V.; {Muriel}, H.; {Ruiz}, A.N.; {Vega-Mart{\'\i}nez}, C.A.; {Cora}, S.A.
\newblock {ROGER: Reconstructing orbits of galaxies in extreme regions using machine learning techniques}.
\newblock {\em Monthly Notices of the Royal Astronomical Society} {\bf 2021}, {\em 500},~1784--1794.
\newblock {\url{https://doi.org/10.1093/mnras/staa3339}}.

\bibitem[{Cora} et~al.(2018){Cora}, {Vega-Mart{\'\i}nez}, {Hough}, {Ruiz}, {Orsi}, {Mu{\~n}oz Arancibia}, {Gargiulo}, {Collacchioni}, {Padilla}, {Gottl{\"o}ber}, and {Yepes}]{sag}
{Cora}, S.A.; {Vega-Mart{\'\i}nez}, C.A.; {Hough}, T.; {Ruiz}, A.N.; {Orsi}, {\'A}.A.; {Mu{\~n}oz Arancibia}, A.M.; {Gargiulo}, I.D.; {Collacchioni}, F.; {Padilla}, N.D.; {Gottl{\"o}ber}, S.;  et~al.
\newblock {Semi-analytic galaxies - I. Synthesis of environmental and star-forming regulation mechanisms}.
\newblock {\em Monthly Notices of the Royal Astronomical Society} {\bf 2018}, {\em 479},~2--24.
\newblock {\url{https://doi.org/10.1093/mnras/sty1131}}.

\bibitem[{Maturi} et~al.(2023){Maturi}, {Finoguenov}, {Lopes}, {Gonz{\'a}lez Delgado}, {Dupke}, {Cypriano}, {Carrasco}, {Diego}, {Penna-Lima}, {Doubrawa}, {V{\'\i}lchez}, {Moscardini}, {Marra}, {Bonoli}, {Rodr{\'\i}guez-Mart{\'\i}n}, {Zitrin}, {M{\'a}rquez}, {Hern{\'a}n-Caballero}, {Jim{\'e}nez-Teja}, {Abramo}, {Alcaniz}, {Benitez}, {Carneiro}, {Cenarro}, {Crist{\'o}bal-Hornillos}, {Ederoclite}, {L{\'o}pez-Sanjuan}, {Mar{\'\i}n-Franch}, {Mendes de Oliveira}, {Moles}, {Sodr{\'e}}, {Taylor}, {Varela}, {V{\'a}zquez Rami{\'o}}, and {Fern{\'a}ndez-Ontiveros}]{2023A&A...678A.145M}
{Maturi}, M.; {Finoguenov}, A.; {Lopes}, P.A.A.; {Gonz{\'a}lez Delgado}, R.M.; {Dupke}, R.A.; {Cypriano}, E.S.; {Carrasco}, E.R.; {Diego}, J.M.; {Penna-Lima}, M.; {Doubrawa}, L.;  et~al.
\newblock {The miniJPAS survey. Cluster and galaxy group detections with AMICO}.
\newblock {\em Astronomy \& Astrophysics} {\bf 2023}, {\em 678},~A145.
\newblock {\url{https://doi.org/10.1051/0004-6361/202245323}}.

\bibitem[{Oman} et~al.(2013){Oman}, {Hudson}, and {Behroozi}]{2013MNRAS.431.2307O}
{Oman}, K.A.; {Hudson}, M.J.; {Behroozi}, P.S.
\newblock {Disentangling satellite galaxy populations using orbit tracking in simulations}.
\newblock {\em Monthly Notices of the Royal Astronomical Society} {\bf 2013}, {\em 431},~2307--2316.
\newblock {\url{https://doi.org/10.1093/mnras/stt328}}.

\bibitem[{Agulli} et~al.(2017){Agulli}, {Aguerri}, {Diaferio}, {Dominguez Palmero}, and {S{\'a}nchez-Janssen}]{2017MNRAS.467.4410A}
{Agulli}, I.; {Aguerri}, J.A.L.; {Diaferio}, A.; {Dominguez Palmero}, L.; {S{\'a}nchez-Janssen}, R.
\newblock {Deep spectroscopy of nearby galaxy clusters - II. The Hercules cluster}.
\newblock {\em Monthly Notices of the Royal Astronomical Society} {\bf 2017}, {\em 467},~4410--4423.
\newblock {\url{https://doi.org/10.1093/mnras/stx371}}.

\bibitem[Cohn(2011)]{10.1111/j.1365-2966.2011.19756.x}
Cohn, J.D.
\newblock Galaxy subgroups in galaxy clusters.
\newblock {\em Monthly Notices of the Royal Astronomical Society} {\bf 2011}, {\em 419},~1017--1027.
\newblock {\url{https://doi.org/10.1111/j.1365-2966.2011.19756.x}}.

\bibitem[Lopes et~al.(2024)Lopes, Ribeiro, and Brambila]{lopes2024role}
Lopes, P.A.; Ribeiro, A.L.; Brambila, D.
\newblock The role of groups in galaxy evolution: compelling evidence of pre-processing out to the turnaround radius of clusters.
\newblock {\em Monthly Notices of the Royal Astronomical Society: Letters} {\bf 2024}, {\em 527},~L19--L25.

\bibitem[Malavasi et~al.(2020)Malavasi, Aghanim, Tanimura, Bonjean, and Douspis]{malavasi2020like}
Malavasi, N.; Aghanim, N.; Tanimura, H.; Bonjean, V.; Douspis, M.
\newblock Like a spider in its web: a study of the large-scale structure around the Coma cluster.
\newblock {\em Astronomy \& Astrophysics} {\bf 2020}, {\em 634},~A30.

\bibitem[Rhee et~al.(2017)Rhee, Smith, Choi, Sukyoung, Jaff{\'e}, Candlish, and S{\'a}nchez-J{\'a}nssen]{rhee2017phase}
Rhee, J.; Smith, R.; Choi, H.; Sukyoung, K.Y.; Jaff{\'e}, Y.; Candlish, G.; S{\'a}nchez-J{\'a}nssen, R.
\newblock Phase-space analysis in the group and cluster environment: time since infall and tidal mass loss.
\newblock {\em The Astrophysical Journal} {\bf 2017}, {\em 843},~128.

\bibitem[{Lisker} et~al.(2013){Lisker}, {Weinmann}, {Janz}, and {Meyer}]{2013MNRAS.432.1162L}
{Lisker}, T.; {Weinmann}, S.M.; {Janz}, J.; {Meyer}, H.T.
\newblock {Dwarf galaxy populations in present-day galaxy clusters - II. The history of early-type and late-type dwarfs}.
\newblock {\em Monthly Notices of the Royal Astronomical Society} {\bf 2013}, {\em 432},~1162--1177.
\newblock {\url{https://doi.org/10.1093/mnras/stt549}}.

\bibitem[Ribeiro et~al.(2013)Ribeiro, de~Carvalho, Trevisan, Capelato, La~Barbera, Lopes, and Schilling]{ribeiro2013spider}
Ribeiro, A.; de~Carvalho, R.; Trevisan, M.; Capelato, H.; La~Barbera, F.; Lopes, P.; Schilling, A.
\newblock SPIDER--IX. Classifying galaxy groups according to their velocity distribution.
\newblock {\em Monthly Notices of the Royal Astronomical Society} {\bf 2013}, {\em 434},~784--795.

\bibitem[{Barrena} et~al.(2024){Barrena}, {Pizzuti}, {Chon}, and {B{\"o}hringer}]{2024A&A...691A.135B}
{Barrena}, R.; {Pizzuti}, L.; {Chon}, G.; {B{\"o}hringer}, H.
\newblock {Unveiling the shape: A multi-wavelength analysis of the galaxy clusters Abell 76 and Abell 1307}.
\newblock {\em Astronomy \& Astrophysics} {\bf 2024}, {\em 691},~A135.
\newblock {\url{https://doi.org/10.1051/0004-6361/202451144}}.

\bibitem[{Vijayaraghavan} et~al.(2015){Vijayaraghavan}, {Gallagher}, and {Ricker}]{2015MNRAS.447.3623V}
{Vijayaraghavan}, R.; {Gallagher}, J.S.; {Ricker}, P.M.
\newblock {The dynamical origin of early-type dwarfs in galaxy clusters: a theoretical investigation}.
\newblock {\em Monthly Notices of the Royal Astronomical Society} {\bf 2015}, {\em 447},~3623--3638.
\newblock {\url{https://doi.org/10.1093/mnras/stu2761}}.

\bibitem[{Einasto} et~al.(2012){Einasto}, {Vennik}, {Nurmi}, {Tempel}, {Ahvensalmi}, {Tago}, {Liivam{\"a}gi}, {Saar}, {Hein{\"a}m{\"a}ki}, {Einasto}, and {Mart{\'\i}nez}]{2012A&A...540A.123E}
{Einasto}, M.; {Vennik}, J.; {Nurmi}, P.; {Tempel}, E.; {Ahvensalmi}, A.; {Tago}, E.; {Liivam{\"a}gi}, L.J.; {Saar}, E.; {Hein{\"a}m{\"a}ki}, P.; {Einasto}, J.;  et~al.
\newblock {Multimodality in galaxy clusters from SDSS DR8: substructure and velocity distribution}.
\newblock {\em Astronomy \& Astrophysics (A\&A)} {\bf 2012}, {\em 540},~A123.
\newblock {\url{https://doi.org/10.1051/0004-6361/201118697}}.

\bibitem[{de Carvalho} et~al.(2017){de Carvalho}, {Ribeiro}, {Stalder}, {Rosa}, {Costa}, and {Moura}]{2017AJ....154...96D}
{de Carvalho}, R.R.; {Ribeiro}, A.L.B.; {Stalder}, D.H.; {Rosa}, R.R.; {Costa}, A.P.; {Moura}, T.C.
\newblock {Investigating the Relation between Galaxy Properties and the Gaussianity of the Velocity Distribution of Groups and Clusters}.
\newblock {\em The Astronomical Journal} {\bf 2017}, {\em 154},~96.
\newblock {\url{https://doi.org/10.3847/1538-3881/aa7f2b}}.

\bibitem[{Healy} et~al.(2021){Healy}, {Blyth}, {Verheijen}, {Hess}, {Serra}, {van der Hulst}, {Jarrett}, {Yim}, and {J{\'o}zsa}]{2021A&A...650A..76H}
{Healy}, J.; {Blyth}, S.L.; {Verheijen}, M.A.W.; {Hess}, K.M.; {Serra}, P.; {van der Hulst}, J.M.; {Jarrett}, T.H.; {Yim}, K.; {J{\'o}zsa}, G.I.G.
\newblock {H I content in Coma cluster substructure}.
\newblock {\em The Astrophysical Journal} {\bf 2021}, {\em 650},~A76.
\newblock {\url{https://doi.org/10.1051/0004-6361/202038738}}.

\bibitem[{Biviano} et~al.(1996){Biviano}, {Durret}, {Gerbal}, {Le Fevre}, {Lobo}, {Mazure}, and {Slezak}]{1996A&A...311...95B}
{Biviano}, A.; {Durret}, F.; {Gerbal}, D.; {Le Fevre}, O.; {Lobo}, C.; {Mazure}, A.; {Slezak}, E.
\newblock {Unveiling hidden structures in the Coma cluster.}
\newblock {\em The Astrophysical Journal} {\bf 1996}, {\em 311},~95--112.
\newblock {\url{https://doi.org/10.48550/arXiv.astro-ph/951211}}.

\bibitem[{Beijersbergen} et~al.(2002){Beijersbergen}, {Hoekstra}, {van Dokkum}, and {van der Hulst}]{2002MNRAS.329..385B}
{Beijersbergen}, M.; {Hoekstra}, H.; {van Dokkum}, P.G.; {van der Hulst}, T.
\newblock {U-, B- and r-band luminosity functions of galaxies in the Coma cluster}.
\newblock {\em Monthly Notices of the Royal Astronomical Society.} {\bf 2002}, {\em 329},~385--397.
\newblock {\url{https://doi.org/10.1046/j.1365-8711.2002.05004.x}}.

\bibitem[{Fukugita} et~al.(1995){Fukugita}, {Shimasaku}, and {Ichikawa}]{1995PASP..107..945F}
{Fukugita}, M.; {Shimasaku}, K.; {Ichikawa}, T.
\newblock {Galaxy Colors in Various Photometric Band Systems}.
\newblock {\em Publications of the Astronomical Society of the Pacific.} {\bf 1995}, {\em 107},~945.
\newblock {\url{https://doi.org/10.1086/133643}}.

\end{thebibliography}




\PublishersNote{}
\end{adjustwidth}

%



\end{document}